\documentclass{aa}  

\usepackage{graphicx}
\usepackage{txfonts}
\begin{document} 

   \title{Origin and evolution of angular momentum of class II disks}

   \author{V.-M. Pelkonen\inst{1,2},
          P. Padoan\inst{1,3},
          M. Juvela\inst{4},
          T. Haugb{\o}lle\inst{5}
          \and
          \AA. Nordlund\inst{5}
          }

   \institute{Institut de Ci\`{e}ncies del Cosmos, Universitat de Barcelona, IEEC-UB, Mart\'{i} i Franqu\'{e}s 1, E08028 Barcelona, Spain\\
              \email{veli.matti.pelkonen@gmail.com}
        \and
              INAF - Istituto di Astrofisica e Planetologia Spaziali, Via Fosso del Cavaliere 100, I-00133 Roma, Italy
         \and Department of Physics and Astronomy, Dartmouth College, 6127 Wilder Laboratory, Hanover, NH 03755, USA
         \and Department of Physics, PO Box 64, 00014 University of Helsinki, Finland
         \and
             Niels Bohr Institute, University of Copenhagen, {\O}ster Voldgade 5-7, DK-1350 Copenhagen, Denmark
             }

   \date{Received September 15, 1996; accepted March 16, 1997}

 
  \abstract
   {While class II pre-main-sequence (PMS) stars have already accreted most of their mass, the continued inflow of fresh material via Bondi-Hoyle accretion acts as an additional mass reservoir for their circumstellar disks. This may explain the observed accretion rates of PMS stars, as well as observational inconsistencies in the mass and angular momentum balance of their disks.}
   {Using a new simulation that reproduces the stellar initial mass function (IMF), we want to quantify the role of Bondi-Hoyle accretion in the formation of class II disks, as well as address the prospect of its observational detection with the James Webb Space Telescope (JWST).}
   {We studied the mass and angular momentum of the accreting gas using passively advected tracer particles in the simulation, and we carried out radiative transfer calculations of near-infrared scattering to generate synthetic JWST observations of Bondi-Hoyle trails of PMS stars.}
   {Gas accreting on class II PMS stars approximately 1\,Myr after their formation has enough mass and angular momentum to strongly affect the evolution of the preexisting disks. The accreted angular momentum is large enough to also explain the observed size of class II disks. The orientation of the angular momentum vector can differ significantly from that of the previously accreted gas, which may result in a significant disk warping or misalignment. We also predict that JWST observations of class II stars will be able to detect Bondi-Hoyle trails with a 80\%-100\% success rate with only a 2\,min exposure time, depending on the filter, if stars with both an accretion rate $\dot{M}> 5 \times 10^{-10}  \, \rm M_\odot/yr$ and a luminosity of $L > 0.5 \, \rm L_\odot$ are selected.}
   {}


\authorrunning{V.-M. Pelkonen et al.} 
\titlerunning{Origin and evolution of angular momentum of class II disks} 
\maketitle
%

\section{Introduction} 
\label{sec:introduction}

Theoretical studies of planet formation assume idealized disk models, in line with a traditional picture of star formation where the stellar mass reservoir is fully contained in a protostellar core: when the collapse of the protostellar core is finished, the circumstellar disk is fully formed and its total mass can only decrease over time. This traditional picture of star formation, known as core collapse, has been recently challenged. Simulations of star formation under physical conditions characteristic of molecular clouds (MCs), such as Larson's relations \citep[e.g.,][]{Larson81,Solomon+87,Heyer+Brunt04}, demonstrate that the timescale of star formation is longer than the free-fall time of prestellar cores, with a significant fraction of the stellar mass coming from larger scales through converging flows \citep{Padoan+20massive,Pelkonen+21,Kuffmeier+23}. Because these converging flows are randomly occurring in supersonic turbulence without the aid of stellar gravity, we refer to them as "inertial inflows" \citep{Padoan+20massive}. The formation time of a star (the duration of the inertial inflows) increases with the final stellar mass, and can exceed $1$\,Myr for massive stars. Being characterized by infall from random dense filaments \citep{Kuffmeier+17,Kuffmeier+18}, this formation phase is quite different from the spherical collapse of a protostellar core, impacting the formation and evolution of protoplanetary disks. 

The same simulations have also shown that once the star has reached its final mass and emerges from the native dense gas, as a pre-main sequence (PMS) class II star, further accretion from larger scales is still possible at a much lower rate through Bondi-Hoyle (BH) accretion. Although too low to increase the mass of the star significantly, early BH accretion may be important for the mass budget of the disk \citep{Padoan+05_BH,Throop+Bally08,Padoan+14_luminosity,Padoan+24_angular}, so it should be included in models of disk evolution. As for the inertial inflows in the formation phase, this mass infall deviates strongly from spherical symmetry, because the relative velocity between the captured gas and the star is much larger than the sound speed, so the accretion proceeds through dense filaments in the BH trail of the star \citep{Padoan+24_angular}. The idea of a very dynamic evolution of disks, characterized by destabilizing filamentary infall well after the initial protostellar collapse, has also been supported by simulations following the early evolution of disks in realistic large-scale environments \citep[e.g.,][]{Kuffmeier+17,Kuffmeier+18}, in idealized setups of late disks \citep{Moeckel+Throop09,Kuffmeier+20,Kuffmeier+21}, and by using large-scale models that capture the infall at late times \citep{Kuffmeier+23,Kuffmeier+24}.

\begin{figure*}
    \centering
    \includegraphics[width=17cm]{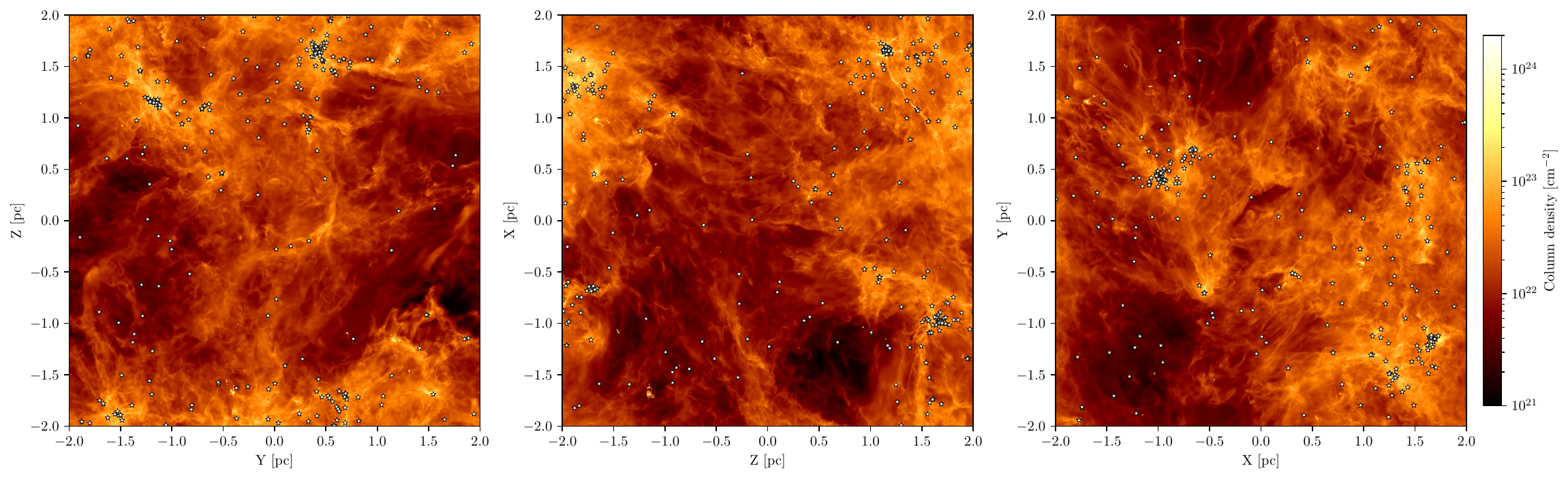}
    \caption{{Column density maps of the simulation from three cardinal directions ($x$, $y$, and $z$). The sink particles are shown with star symbols.}}
    \label{fig_column_density}
\end{figure*}

\citet{Padoan+24_angular} have shown that BH infall in class II stars can account not only for the observed disk masses, but also for the disk specific angular momentum, or disk sizes, which implies that BH infall, rather than the initial protostellar collapse, may be the main formation mechanism of class II disks. They found that, in supersonic turbulence, the specific angular momentum increases linearly with scale, $j(R)\sim R$, and they derived the angular momentum of gas captured by BH infall, $j_{\rm BH}$, as the characteristic value of $j(R)$ at a scale equal to the BH radius, $R_{\rm BH}$, of the PMS star, $j_{\rm BH}=j(R_{\rm BH})$, using the following expression for the BH radius:
\begin{equation}
R_{\rm BH}=\frac{2 G M_{\rm star}}{c_{\rm s}^2+v_{\rm rel}^2}, 
\label{R_BH}
\end{equation}
where $c_{\rm s}$ is the sound speed and $v_{\rm rel}$ the relative velocity between the gas and the star. The expression reduces to the Hoyle-Lyttleton radius, $R_{\rm HL}=2 G M_{\rm star}/v_{\rm rel}^2$, in the pressureless case where $c_{\rm s}=0$ \citep{Hoyle+Lyttleton39}, and to the Bondi radius for spherical accretion, $R_{\rm B}=G M_{\rm star}/c_{\rm s}^2$, in the limit of $v_{\rm rel} = c_{\rm s}$ \citep{Bondi52}.
The characteristic specific angular momentum of the gas captured by the star at the BH radius is
\begin{equation}
j_{\rm BH} = 4.3\times10^{20} {\rm cm}^2\,{\rm s}^{-1}\, (\sigma_{\rm v,rel}/2 \,{\rm km\,s^{-1}})^{-1}\,(M_{\rm star}/1\,M_\odot)
\label{j_BH_t_main}
\end{equation}
and the corresponding disk radius (assuming a simple Keplerian disk model),
\begin{equation}
R_{\rm d} = 3.6\times10^2 {\rm au} \,(\sigma_{\rm v,rel}/2 {\rm \,km\,s^{-1}})^{-2}\,(M_{\rm star}/1\,M_\odot)
= R_{\rm BH}/4.1
\label{R_d_BH_t_main}
\end{equation}
where $\sigma_{\rm v,rel}$, the standard deviation of the gas velocity relative to the star's velocity, has been defined to include the sound speed, $\sigma_{\rm v,rel}^2 = {c_{\rm s}^2+v_{\rm rel}^2}$, and we have adopted the same normalization of $2\,{\rm \,km\,s^{-1}}$ as in \citet{Padoan+24_angular}, because it is a typical value for the stars in the simulation in the early BH phase.

Eqs.\,(\ref{j_BH_t_main}) and (\ref{R_d_BH_t_main}) should be considered as upper limits for the average values of $j_{\rm BH}$ and $R_{\rm d}$, because the angular momentum of the gas captured by a star in its trajectory is generally not constant over a $\sim 1$\,Myr timescale, causing partial cancelation in the average angular momentum vector, as mass accumulates in the disk. In addition, the scatter in the turbulence scaling of $j(R)$ and $\sigma_{v}(R)$ is expected to cause a significant scatter of $M_{\rm d}$, $j_{\rm BH}$, and $R_{\rm d}$ around the mean values predicted by the above relations. The time evolution of the angular momentum of the accreting gas was not addressed in \citep{Padoan+24_angular}, because of the Eulerian nature of the numerical analysis, where the angular momentum was computed within a sphere of radius equal to $R_{\rm BH}$ around each star. In this work, we continue the analysis of the same simulation, but with a Lagrangian perspective, using tracer particles to track the time evolution of the angular momentum of the gas accreted by the PMS stars in the simulation. We also address the potential for observational studies of BH infall in PMS stars, by carrying out synthetic James Webb Space Telescope (JWST) observations of BH trails from the simulation.  

The paper is structured as follows. In \S\,2 we briefly describe the simulation and the BH infall rates in comparison with observed accretion rates of PMS stars (\S\,2.1); we also present the method of tracking the angular momentum evolution using the tracer particles in the simulation (\S\,2.1). Our results on the angular momentum of the gas captured by BH infall and its time evolution are presented in \S\,3. In \S\,4 we describe the synthetic near-infrared (NIR) scattering observations around selected PMS stars from the simulation. Results are further discussed in \S\,5 in the context of actual observations of streamers, and our conclusions are summarized in \S\,6.

\section{Numerical methods}

\subsection{Simulation and accretion rates}  \label{sec:simulation} 

This work is based on the analysis of a numerical simulation of randomly driven, supersonic, magneto-hydrodynamic (MHD) turbulence, including self-gravity and sink particles to capture the star-formation process. It is the same global simulation used in \citet{Jorgensen+22}, \citet{Jensen+23}, \citet{Kuffmeier+23,Kuffmeier+24}, and \citet{Padoan+24_angular}, and it is equivalent to the ph{high} reference simulation in \citet{Haugbolle+18imf}, except that the numerical resolution (the root grid) is larger by a factor of two, corresponding to a factor of 8 increase in the number of computational cells. Details of the numerical method can be found in \citet{Haugbolle+18imf} and are only briefly summarized here. The simulation solves the three-dimensional MHD equations with the adaptive-mesh-refinement (AMR) code RAMSES \citep{Teyssier07}, with a root grid of $512^3$ cells, representing a region with a size of 4\,pc, and six levels of refinement, corresponding to a smallest cell size $\Delta x=25$\,au. The total mass, mean density, and mean magnetic field strength are $M_{\rm box}=3000 \rm \,M_\odot$, ${\bar n_{\rm H}}=1897 \, \rm cm^{-3}$, and ${\bar B}=7.2 \, \mu \rm G$. We use an isothermal equation of state with a gas temperature $T=10$\,K and periodic boundary conditions. The turbulence is first driven, without self-gravity, for $\approx 20$ dynamical times, with a random solenoidal acceleration giving an rms sonic Mach number of approximately 10. The simulation is then continued for $\sim 2$\,Myr (snapshot number 448) with self-gravity and sink particles (maintaining the random driving force) to capture the ongoing formation of individual stars, yielding 317 stars with a mass distribution consistent with the observed stellar initial mass function (IMF) \citep{Salpeter55,Chabrier05}. Figure~\ref{fig_column_density} shows the column density maps of the simulation from three lines-of sight, with the stars overplotted on the maps. 

\begin{figure}
    \centering
    \includegraphics[width=\hsize]{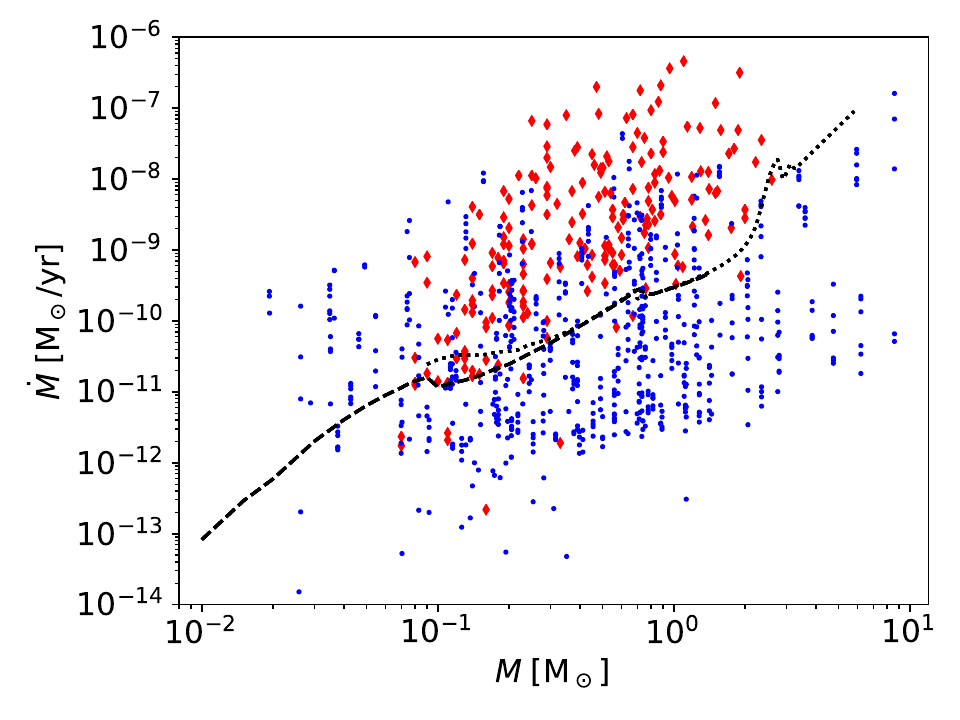}
    \caption{{Mass accretion rates as a function of stellar mass. Red diamonds are Class~II YSOs in Chamaeleon I, Lupus, and L1688 from \citet{Testi+22}, and in Taurus from \citet{Gangi+22}. Blue points are large-scale infall rates of star particles from the MHD simulation (see Section\,\ref{sec:simulation}) with local gas density lower than $5\times 10^4$\,cm$^{-3}$. The dashed and dotted black lines are the expected chromospheric noise levels \citep{Manara+13,Manara+17} using the 3\,Myr isochrones of \citet{Baraffe+15} and \citet{Feiden+16}, respectively.}} 
    \label{fig_accretion}
\end{figure}

The spatial resolution of the simulation is insufficient to resolve the accretion disks around the sink particles, and the mass and angular momentum loss by winds and jets. Thus, the sink particle should be viewed as containing both the star and its accretion disk. To mimic the mass loss by protostellar winds and jets, the sub-grid model for mass accretion onto the sink particles \citep[see][for details]{Haugbolle+18imf} prescribes that only 50\% of the gravitationally captured mass (the infall onto the disk) is accreted onto the sink particles; the other 50\% of the captured mass is removed from the simulation. Although we use the terms "accretion" and "infall" interchangeably, with regard to the simulation, we always refer to the mass infalling toward the (unresolved) disks, but whenever we give a numerical value for the infall (e.g., in Fig.\,\ref{fig_accretion}), it is already halved to account for the mass loss. We assume that the infalling mass will settle onto a disk according to its angular momentum, but the accretion from the disk to the star (i.e., the depletion of the disk mass), which is measured by the observers, is not captured by the simulation. However, by comparing the computed infall rates (from the interstellar medium (ISM) to the disks) with the observed accretion rates (from the disks to the stars) of real stars we can evaluate the importance of the infall for the mass budget of disks.

Fig.~\ref{fig_accretion} shows the infall rates, $\dot {M}$, versus the stellar mass, $M$, for a subset of the stars found in the final six snapshots of the simulation (a period of 30\,kyr). Because this work and the observational samples we compare with focus on class II stars, we have selected for this figure the sink particles that are representative of class II stars according to the criterion discussed in \citet{Padoan+24_angular}, that is the gas density at the sink particle location must be
$\le 5\times 10^4$\,cm$^{-3}$. This criterion yields a total of 672 stars (from the six snapshots), with an upper limit on the accretion rate of order $10^{-7}\,M_{\odot}\,{\rm yr}^{-1}$. The median and maximum ages of the stars are 1.1 and 2.0\,Myr respectively, consistent with estimated ages of class II stars \citep{Evans+09}. Fig.~\ref{fig_accretion} also shows the accretion rates of 207 class II stars in four star-forming regions with estimated ages $\lesssim 3$\,Myr \citep{Testi+22,Gangi+22}. The infall rates from the simulation are comparable to the observed accretion rates, except for the lack of infall rates above $\sim 10^{-7}\,M_{\odot}\,{\rm yr}^{-1}$, due to our conservative classification of class II sink particles, and the lack of low accretion rates in the observations, because of the sensitivity limit arising from chromospheric noise in the spectra \citep{Manara+13,Manara+17}. As the computed infall rates onto the disks are comparable to the observed accretion rates from the disks to the stars, this late infall must play an important role in the disk mass budget.

\subsection{Tracer particles and angular momentum evolution}
\label{S_tracers}

The simulation includes $512^3$ passively advected tracer particles, each representing 
a fluid element with a characteristic mass of approximately $2.2\times 10^{-5}\,M_{\odot}$. The tracer particles are advected with a symplectic kick-drift-kick scheme, using the fluid velocities sampled with cloud-in-cell interpolation. They record all the hydrodynamic variables and are tagged once they accrete onto a sink particle. By selecting all the tracer particles that eventually accrete onto a given star, we can track the Lagrangian history of fluid elements contributing to the mass of that star, with a time resolution of 5\,kyr, that is the time interval between the simulation snapshots that were saved. Because observed disk masses in class II stars are of order 1\% of the stellar mass on average \citep{Testi+22}, we study the time evolution of the accreting gas in the last ten epochs, where each epoch corresponds to the accretion of 1\% of the stellar mass, as representative of the accretion process that is most likely responsible for the current dynamical state of class II disks. 

\begin{figure}
    \centering
    \includegraphics[width=\hsize]{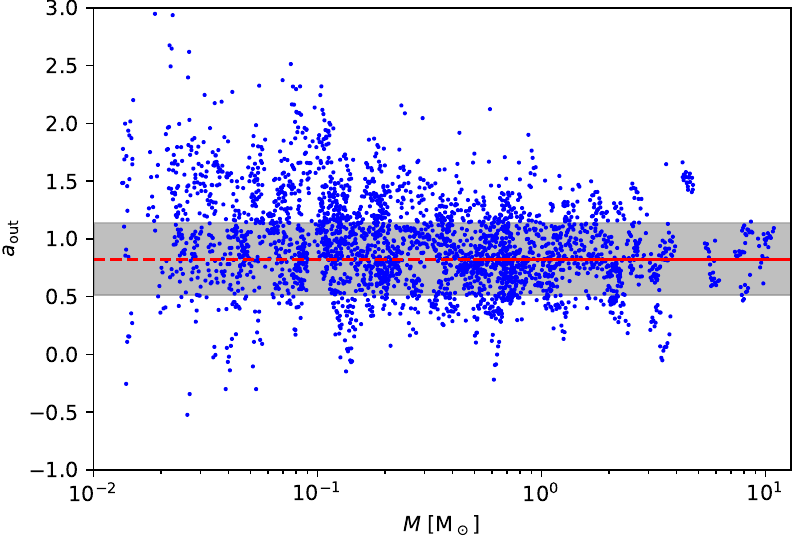}  
    \includegraphics[width=\hsize]{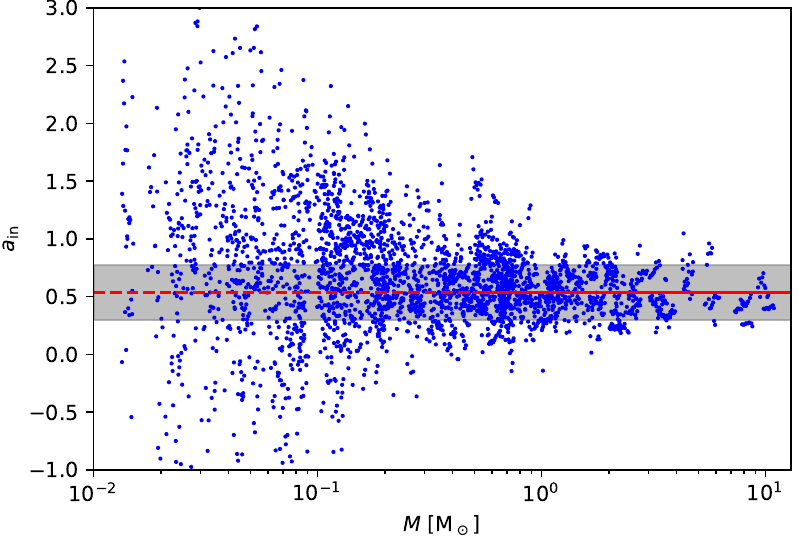}  
    \caption{{Power-law slope of $j_{\rm tr}$ versus $r_{\rm tr}$, $j_{\rm tr} \propto r_{\rm tr}^{\, a_{\rm out,in}}$, averaged for all the tracer particles within a single epoch, versus the final stellar mass (blue dots), as explained in the main text. The upper panel is for the scaling outside of $R_{\rm BH}$, the lower panel for the scaling inside $R_{\rm BH}$. The red lines show the median slope values $a_{\rm out}= 0.82$ (upper panel) and $a_{\rm in}= 0.53$ (lower panel, computed over masses $\ge 0.5\,\rm M_\odot$), and the shaded regions the rms value around the median.}}
    \label{fig_tr_slope}
\end{figure}

Although the disk is not resolved in the simulation, we aim at measuring the angular momentum supplied to the disk by the accreting gas. To that purpose, following the reasoning in \citet{Padoan+24_angular}, which is summarized in \S\,\ref{sec:introduction}, we assume that the disk receives the angular momentum of the gas gravitationally captured by the star, hence the angular momentum is measured at a scale equal to the Bondi-Hoyle radius of the star, $R_{\rm BH}$. 
This is done by computing the angular momentum vector of each accreting tracer particle, $\boldsymbol{J}_{\rm tr}$, at a distance $R_{\rm BH}$ from the star: 
\begin{equation}\label{eq:jtr}
  {\boldsymbol{J}}_{{\rm tr}} = m_{\rm tr} \; {\boldsymbol{r}}_{\rm tr} \times {\boldsymbol{v}}_{\rm tr},
\end{equation}
where $m_{\rm tr}$ is the mass of the individual tracer particle, and ${\boldsymbol{r}}_{\rm tr}$ and ${\boldsymbol{v}}_{\rm tr}$ are the position and velocity vectors of the individual tracer particle with respect to the accreting star when the tracer particle crosses $R_{\rm BH}$. Due to the tracer particle data being recorded every 5\,kyr, the last position of a tracer is often outside (low-mass stars) or inside (high-mass stars) of $R_{\rm BH}$. Thus, we need to interpolate or extrapolate $\boldsymbol{J}_{\rm tr}$ at the distance $R_{\rm BH}$ based on its values at the available distances. For that purpose, we examine the dependence of the specific angular momentum of each tracer particle, $j_{\rm tr}$, on the distance, $r_{\rm tr}$, between the tracer and the star accreting it. We expect that $j_{\rm tr}$ scales with $r_{\rm tr}$ according to different laws outside and inside $R_{\rm BH}$, hence we consider the two cases separately, as explained in the following paragraph.

We first take all the tracers contributing to the last 10\% of the mass of each star, splitting them into ten 1\% mass bins (epochs) in chronological order of their accretion time. The chronological order among the tracers accreted in the same snapshot is not available, so such tracers are sorted by their index number, which, by this point in the simulation, is effectively a random order. For every tracer particle in each epoch and each star, we take the tracer's last ten positions (50\,kyr) before crossing $R_{\rm BH}$ (when the tracer is outside $R_{\rm BH}$), and derive the least-square fit slope of $j_{\rm tr}$ versus $r$ in log-log space, $j_{\rm tr} \propto r_{\rm tr}^{\, a_{\rm out,in}}$. We then compute the mean value of the slope, $a_{\rm out}$, by averaging the slopes of all the tracer particles in each mass bin of each star. The procedure is then repeated for the 10 positions after the tracer particle has crossed $R_{\rm BH}$ (when the tracer is inside $R_{\rm BH}$), yielding the mean slopes $a_{\rm in}$. It is possible that the tracer particles spend less than 50\,kyr between crossing $R_{\rm BH}$ and accreting onto the star, in which case the least-square fit to derive the slope relies on less than 10 positions. When less than three positions are available, the slope is not derived for that tracer particle. If no slope is derived for any of the tracers within a single epoch (1\% mass bin) of a specific star, no $a_{\rm in}$ value is derived for that epoch. This happens only for low-mass stars. Moreover, in the case of low-mass stars, the value of $R_{\rm BH}$ is usually low, corresponding to scales that are not well resolved in the simulation. Hence, we prefer not to rely on the values of $a_{\rm in}$ derived for low-mass stars, but only on those for stars with mass $\ge 0.5$\,M$_{\odot}$.

\begin{figure*}
    \centering
    \includegraphics[width=16cm]{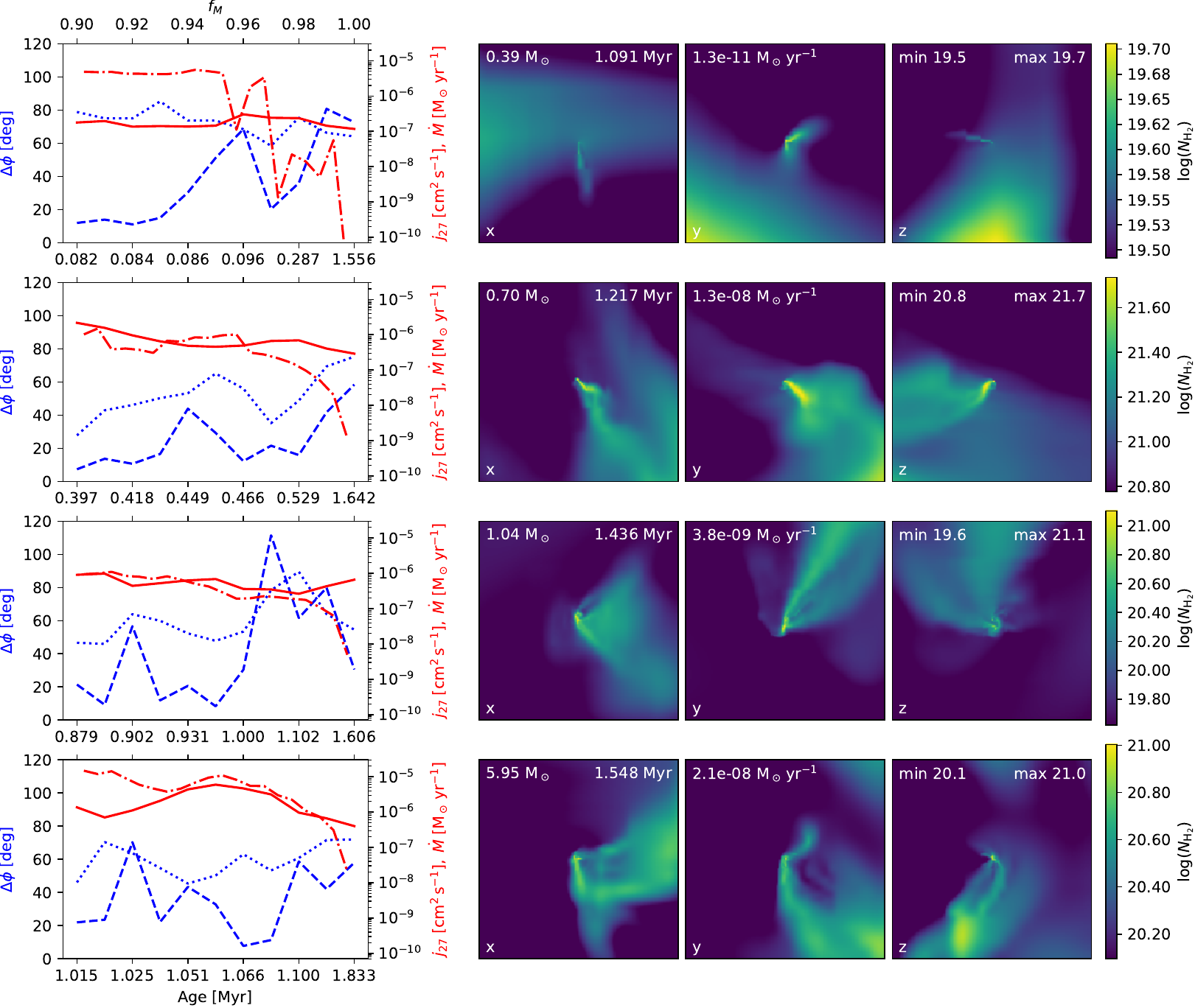}
    \caption{{Four example stars (one per row) showing the accretion of the last 10\% of their mass on the left (the top x-axis is the mass fraction, $f_M$, and the bottom x-axis is the corresponding age of the star), and their late-accretion flows on the right. Left column: Angle $\Delta \phi$ (dashed blue line, "disk-to-disk" misalignment), angle $\delta \phi$ (dotted blue line, "tracers-to-disk" dispersion), specific angular momentum of the epoch (solid red line), $j_{27} = {j}\times{10^{-27}}$, and the median accretion rate within 0.5\% massbins (dash-dotted red line), $\dot{M}$. Right column: Column density images of $log (N_{\rm H_2})$ in a 6400 au box, from three different cardinal directions (xyz), using the middle snapshot of the last epoch for each of the stars. The text in the images list the final mass, the age and the average accretion rate at this snapshot, and the minimum and the maximum limits used for the colorscale of the logarithm images, in all directions.}}
    \label{fig_example}
\end{figure*}

Figure\,\ref{fig_tr_slope} shows the values of $a_{\rm out}$ (upper panel) and $a_{\rm in}$ (lower panel) as a function of the star's final mass. The horizontal red lines correspond to the median values of the slopes, measured over the whole mass range for $a_{\rm out}$, and for $M\ge 0.5$\,M$_{\odot}$ in the case of $a_{\rm in}$, as explained above. These median values are $a_{\rm out}=0.82$ and $a_{\rm in}=0.53$. Interestingly, the median value of $a_{\rm in}$ shows that the tracer particles inside $R_{\rm BH}$ have velocities close to Keplerian, which is consistent with the prediction that the disk size should be just a few times smaller than $R_{\rm BH}$ (see Eq.\,(\ref{R_d_BH_t_main})).
We apply these median scaling exponents to rescale the angular momentum of each tracer particle to a distance equal to $R_{\rm BH}$. For each tracer with at least one recorded position inside $R_{\rm BH}$, we scale its first inner $\boldsymbol{J}_{\rm tr}$ value back to $R_{\rm BH}$ using the $a_{\rm in}$ slope. If no position inside $R_{\rm BH}$ is recorded, we take the tracer's last position before accretion and scale its $\boldsymbol{J}_{\rm tr}$ value to $R_{\rm BH}$ using the median slope $a_{\rm out}$. Finally, we scale the angular momentum of each tracer to the distance of the expected disk radius from Eq.\,(\ref{R_d_BH_t_main}), $R_{\rm d}=R_{\rm BH}/4.1$, using the median slope $a_{\rm in}$. All rescaling applies to the modulus of $\boldsymbol{J}_{\rm tr}$, the direction is left unchanged. The scaling procedure is summarized in the following formulas, 
\begin{equation}\label{eq:jtrdisk}
  {\boldsymbol{J}}_{{\rm tr,disk}} =
    \begin{cases}
      {\boldsymbol{J}}_{\rm tr} \times (R_{\rm BH}/r_{\rm tr})^{a_{\rm in}} / \, 4.1^{a_{\rm in}} & \text{, if $r_{\rm tr} < R_{\rm BH}$,}\\
      {\boldsymbol{J}}_{\rm tr} \times (R_{\rm BH}/r_{\rm tr})^{a_{\rm out}} / \, 4.1^{a_{\rm in}} & \text{, if $r_{\rm tr} \ge R_{\rm BH}$,}\\
    \end{cases}       
\end{equation}
where ${\boldsymbol{J}}_{{\rm tr,disk}}$ is the rescaled angular momentum vector of an individual tracer particle at the distance of the expected outer edge of the disk.

\begin{figure*}
    \centering
    \includegraphics[width=16cm]{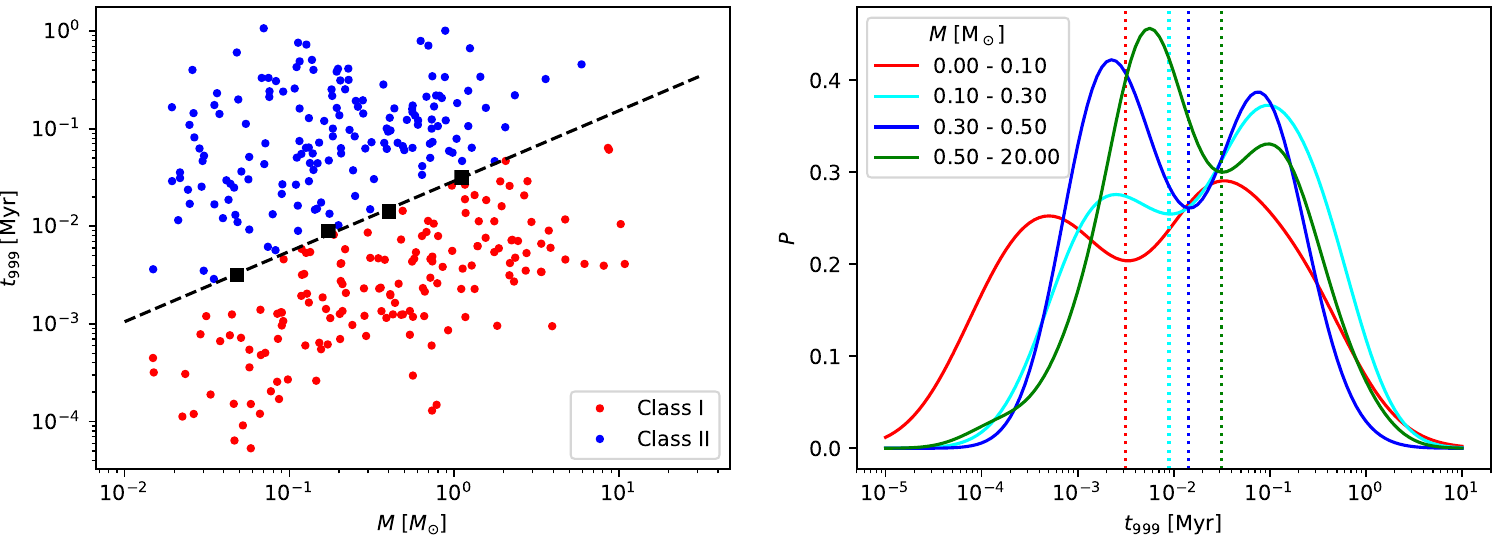}
    \caption{{Scatter plot of the formation time of 90\% of the last epoch (t999) versus mass on the left, and kernel density estimator (KDE) probability density plots of the formation time in four bins of final stellar mass on the right. We find the local minima of the KDE functions, which are then plotted on the scatter plot as black squares. The black dashed line is a fit to the black squares, $\log \, y = 0.72 \pm 0.03 \, \log \,x - 1.54 \pm 0.03$. This line is used to separate the sinks fed by slower Bondi-Hoyle accretion (blue points) and sinks accreting at a much higher rate, likely still in the class I or even earlier protostellar phases (red points).}}
    \label{fig_time}
\end{figure*}

The total angular momentum vector, $\boldsymbol{J}$, in each epoch of each star, is then calculated by summing over the angular momentum vectors of the tracers accreted during that epoch:
\begin{equation}
  {\boldsymbol{J}} = \sum_{i=0}^{n}{\boldsymbol{J}}_{{\rm tr,disk,i}}.
\end{equation}
The corresponding specific angular momentum is derived from the length of the angular momentum vector divided by the mass accreted during that epoch (1\% of the final stellar mass):
\begin{equation}
  {j} = \frac{|{\boldsymbol{J}}|}{\sum_{i=0}^{n}m_{\rm tr,i}},
\end{equation}
where $m_{\rm tr,i}$ is the mass of the individual tracer particles that are accreted during that particular epoch.

In addition, we calculate the mass-weighted average angle, $\delta \phi$, between the angular momentum vector of individual mass tracer particles, ${\boldsymbol{J}}_{{\rm tr,disk,i}}$, and their summed up angular momentum vector of each epoch, ${\boldsymbol{J}}$. This is a measure of how ordered the accretion was within an epoch ("tracers-to-disk" dispersion). We also calculate the angle $\Delta \phi$ between the angular momentum vectors of consecutive epochs (${\boldsymbol{J}_{\rm k}}$ vs. ${\boldsymbol{J}_{\rm k-1}}$), measuring the misalignment between the epochs ("disk-to-disk" misalignment). We use the following equations:
\begin{equation}
  \delta \phi = \frac{\sum_{i=0}^{n} m_{\rm tr,i} \, \arccos (\frac{{\boldsymbol{J}}_{{\rm tr,disk,i}} \cdot {\boldsymbol{J}}}{|{\boldsymbol{J}}_{{\rm tr,disk,i}}| \, |{\boldsymbol{J}}|})}  {\sum_{i=0}^{n}m_{\rm tr,i}},
  \label{eq:delta_phi}
\end{equation}
and
\begin{equation}
  \Delta \phi = \arccos (\frac{{\boldsymbol{J}_{\rm k}} \cdot {\boldsymbol{J}_{\rm k-1}}}{|{\boldsymbol{J}_{\rm k}}| \, |{\boldsymbol{J}_{\rm k-1}}|}).
  \label{eq:Delta_phi}
\end{equation}

Figure\,\ref{fig_example} shows four example stars and the accretion of the last 10\% of their mass, as well as local column density images of the middle point in time of the accretion of the last epoch (the last 1\% of the accreted mass), which clearly shows the BH accretion wakes. They all share some general characteristics: the accretion rate starts high and then drops quite rapidly toward the end, as the star is no longer accreting from converging flows and instead accretes via BH accretion. The mass of the star does seem to influence in which epoch this switch happens. The least massive star (top-most) accretes very quickly at first, almost 1\% of its mass per kyr, before its accretion rate quickly drops around 97\% of its final mass. In the case of the second-least massive star, this transition is more gradual, and the last two epochs are accreted significantly more slowly. By contrast, in the two most massive stars, the accretion drops significantly only toward the last epoch. Since all stars that are fully formed in the simulation have both the accretion rate and the flow morphology consistent with BH accretion in the epoch when they receive the last 1\% of their mass, we will use this last epoch (the final mass bin plotted in Figure\,\ref{fig_example}) to study the contribution of BH accretion to the disk formation process.

In addition to the accretion rate, Figure\,\ref{fig_example} also shows the evolution of the magnitude of the specific angular momentum, $j$, and of the two angles $\delta \phi$ and $\Delta \phi$ defined above. There is a general trend of $j$ to slightly decrease toward the last epoch, as might be expected when the stars decouple from their surrounding gas, causing a reduction of their Bondi-Hoyle radius (from the increased relative velocity between the star and the gas), with an exception to that trend shown in the third panel from the top. It is also notable that the angle $\delta \phi$ is rather large, indicating the accretion of material with different angular-momentum directions, and that the angle $\Delta \phi$ is occasionally very large, signaling episodes of sudden change in the average direction of the accreted angular momentum, which is discussed further in \S\,\ref{sec:angular.2}.

\section{Disk angular momentum and alignment} \label{sec:angular}

\subsection{Extraction of class II sources} \label{sec:angular.1}

Because we seek to characterize the contribution to the disk formation process by BH accretion in class II PMS stars, we need to target our analysis to stars that reach the class II phase before the end of the simulation. A simple way to identify stars that have reached nearly 100\% of their real final mass before the end of the simulation is to consider their average accretion rate during the time they accumulate the last 1\% of their total mass at the end of the simulation. If the accretion rate is low enough, the star is essentially fully formed, though the remaining BH accretion may still be significant for the mass balance of the disk. The value of the accretion rate that separates the formation phase from the BH phase (essentially class I or earlier from class II), is expected to be mass dependent, as we have previously demonstrated that the formation time (for example to reach 95\% of the final mass), scales approximately as the square root of the final stellar mass, $\sim M^{1/2}$ \citep{Padoan+14_luminosity,Haugbolle+18imf,Padoan+20massive,Pelkonen+21}, hence the average accretion rate scales approximately as $\sim M^{1/2}$ (though with a large scatter, increasing toward lower masses). 

We express the accretion rate in terms of the time needed to accrete the last 1\% of the stellar mass, $t_{999}$\footnote{The subscript reflects the fact that we actually use the interval between 99\% and 99.9\% of the mass at the end of the simulation.}, which is plotted as a function of the stellar mass at the end of the simulation in the left panel of Figure\,\ref{fig_time}. As apparent from the distribution of the points in the scatter plot, and confirmed by the histograms in the right panel of the same figure, the stars are separated into two populations, where the transition value of $t_{999}$ has a mass dependence, $t_{999}\sim M^{0.7}$, qualitatively consistent with the above argument based on the star-formation time (which would predict $t_{999}\sim M^{0.5}$). Because the transition between the two populations is well-defined in the scatter plot, we identify as class I (or earlier embedded phases) the stars with $t_{999}$ values below the transition (red dots in the left panel of Figure\,\ref{fig_time}), and as class II all the stars above the transition (blue dots in the figure). This selection returns 160 class II stars, out of a total of 317 stars. The fraction of class II stars is lower than that of nearby star-forming regions because the median age of the stars is only 0.79\,Myr (less than half the duration of the simulation with self-gravity, 2\,Myr, as the star-formation rate increases with time).

\subsection{Specific angular momentum and alignment of class II disks} \label{sec:angular.2}

\begin{figure}
    \centering  
    \includegraphics[width=\hsize]{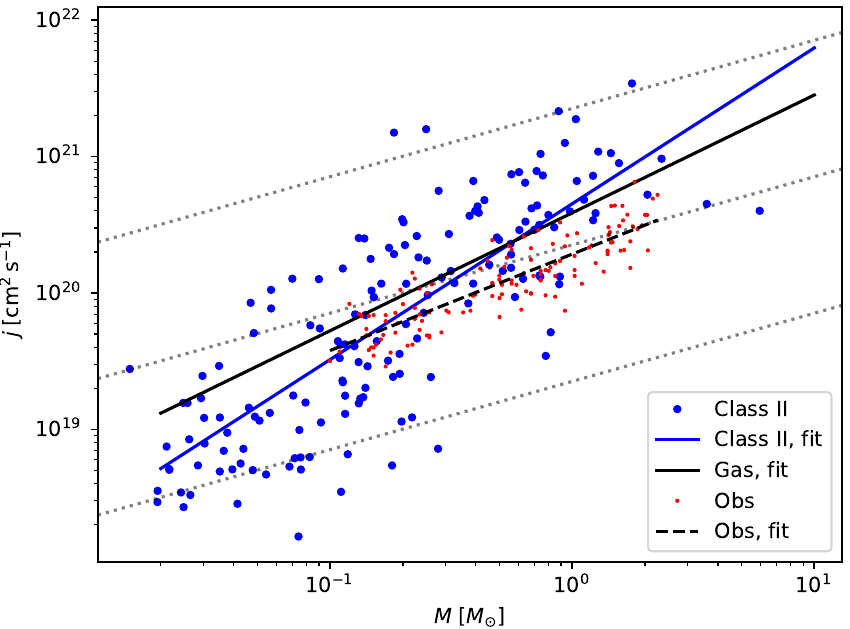}    
    \caption{{Scatter plot of specific angular momentum versus mass in the last epoch of the class II stars in the simulation (blue dots), and for observations of spatially resolved disks \citep{Hendler+20,Cieza+21,Long+22,Stapper+22} (smaller red dots). The blue line is a least-squares fit ($\log y = (1.14\pm0.07) \, \log x + (20.7\pm0.06$)) to the blue data points, while the solid black line is the fit ($\log y = (0.86\pm0.03) \, \log x + (20.6\pm0.03)$) from the Eulerian analysis in \citet{Padoan+24_angular} and the dashed black line a fit ($\log y = (0.71\pm0.04) \, \log x + (20.3\pm0.02$)) to the observations. The gray dotted lines are the specific angular momentum for a given disk radius (1, 100, and 10000 au, with increasing $j$) as a function of stellar mass, using Eq. 29 in \citet{Padoan+24_angular} for Keplerian disks.}}
    \label{fig_j}
\end{figure}

For each class II star in the simulation, we compute the average specific angular momentum, $\boldsymbol j$, of the last ten 1\% mass bins (epochs) as explained in \S\,\ref{S_tracers}. We consider the last epoch as the most representative contribution of BH accretion to the formation of class II disks, and plot the absolute value of its specific angular momentum vector as a function of the final stellar mass, $M$, in Figure\,\ref{fig_j} (blue dots). There is a clear positive correlation between $j$ and $M$, with a least-squares fit giving the relation:
\begin{equation}
j = 4\times10^{20} {\rm cm}^2\,{\rm s}^{-1}\, (M/1\,M_\odot)^{1.1}. 
\label{eq_j_M}
\end{equation}
This result is in striking agreement with the analytical prediction (Eq.\,(\ref{j_BH_t_main})), as well as the result of the Eulerian analysis in our previous paper \citep{Padoan+24_angular}, which is shown by the black line in Figure\,\ref{fig_j}. We also find a significant overlap between our predicted $j$ values and those derived from observations of disk sizes \citep{Hendler+20,Cieza+21,Long+22,Stapper+22}, shown by the smaller red dots in Figure\,\ref{fig_j}. Our predicted $j$ values are on average a factor of two higher than those of the observed disks, which is to be expected because there are multiple processes that could contribute to angular momentum transport when the actual disk is assembled from the gravitationally captured gas. Such processes are not included here, as the simulation does not resolve the disk formation, and we predict the maximum angular momentum by sampling it at the disk edge. 

We also performed the above study using disk mass proportions that scaled with the square root of the stellar mass: $m_{\rm d}/m_{\rm s} = 0.01 \times (m_{\rm s}/1\, \rm M_\odot)^{0.5}$, motivated by \citet{Testi+22}. The results were very similar to the 1\% case, but with lower statistics for the low-mass stars which did not have enough mass tracer particles.

\begin{figure}
    \centering
    \includegraphics[width=\hsize]{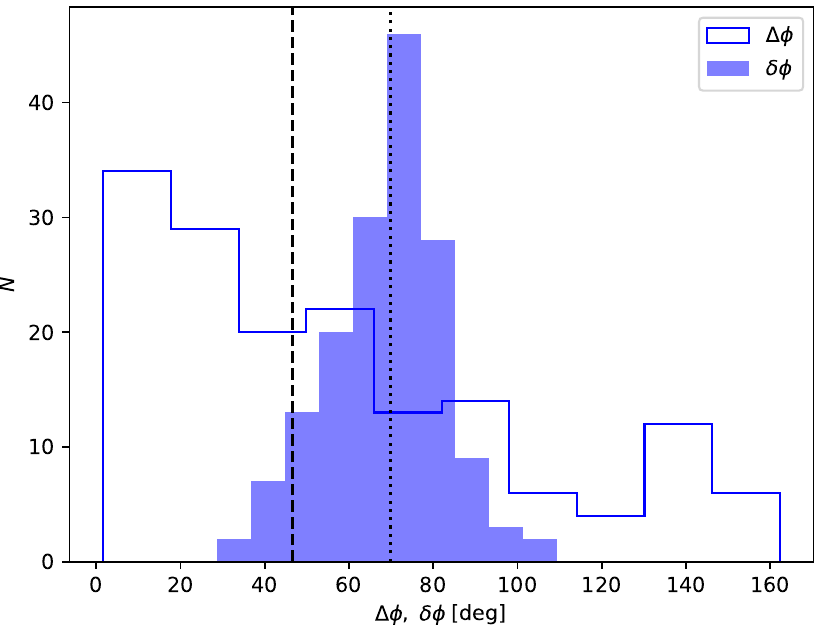}
    \caption{{Histograms of the last $\Delta \phi$ angle (blue step histogram), and of the last $\delta \phi$ angle (blue-shaded bar histogram), for the class II stars in the simulation (see the main text for the definition of the angles). The median values of $\Delta \phi$ and $\delta \phi$ are shown by the black dashed and dotted vertical lines, respectively.}}
    \label{fig_angle_hist}
\end{figure}

An important consequence of BH accretion in class II disks is the possibility of disk misalignment. The mass accretion in the BH phase is a stochastic process that depends on the gas mass distribution encountered by the class II star in its trajectory within the parent cloud. Both the mass accretion rate and the accreted angular momentum can have significant time fluctuations. The angular momentum arises primarily from the relative velocity between the gas and the star, and the offset between the center of mass of the captured gas and the position of the star, rather than from local gas vorticity \citep[see discussion in][]{Padoan+24_angular}. Thus, the direction of the angular momentum vector of the accreted gas can change over time. To quantify this effect in the simulation, we compute the angle between the average $\boldsymbol J$ vectors in the last two epochs of every class II star, as defined in Eq.\,(\ref{eq:Delta_phi}). The result is shown as a histogram in Figure\,\ref{fig_angle_hist}. The $\Delta \phi$ angle distribution peaks at low values, but cover a broad range from $0^{\circ}$ to $160^{\circ}$, with a median value of 
$47^{\circ}$. This suggests that, besides a significant contribution to the disk mass budget,  BH accretion in class II stars may also cause disk misalignment or warping, and strongly affect the angular momentum evolution of the disks.         
Thanks to the tracer particles, besides the change in direction of the mean $\boldsymbol J$ vector between different epochs, $\Delta \phi$, we can also compute the internal dispersion in the direction of the angular momentum of each tracer particle contributing to a single epoch (a 1\% mass bin), $\delta \phi$, as defined in Eq.\,(\ref{eq:delta_phi}). The probability distribution of $\delta \phi$ is shown by the shaded histogram in Figure\,\ref{fig_angle_hist}, and has a median value of $70^{\circ}$. This corresponds to a cancelation factor of approximately three for the modulus of $\boldsymbol J$ obtained from the vector sum of individual tracer particles. So, the cancelation effect is significant, but the accreted angular momentum is far from being completely random to the point of resulting in nearly total cancelation. This relatively modest cancelation effect explains the good agreement of our result, expressed by Eq.\,(\ref{eq_j_M}), with the analytical prediction from Eq.\,(\ref{j_BH_t_main}) and the Eulerian analysis in \citet{Padoan+24_angular}.

\subsection{Embedded phase versus class II}

\begin{figure}
    \centering  
    \includegraphics[width=\hsize]{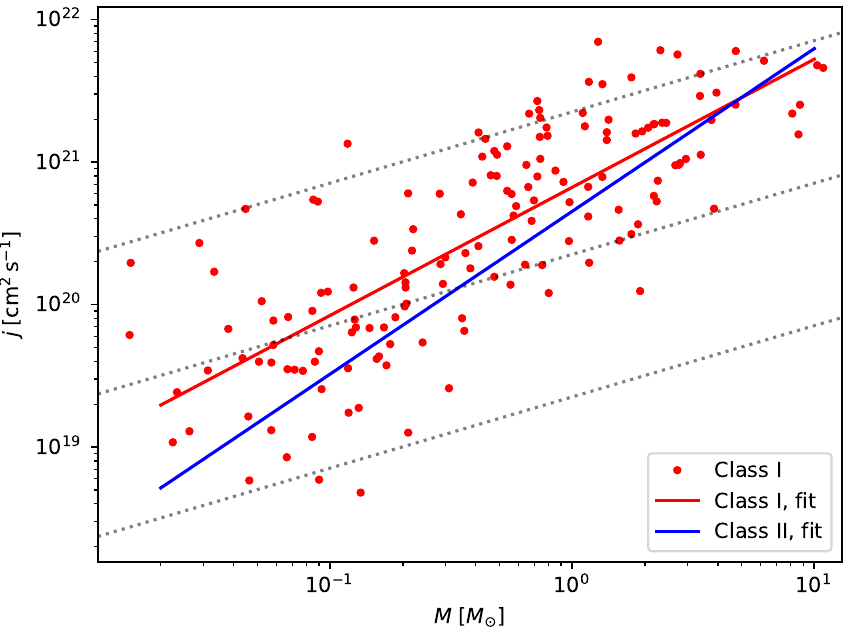}    
    \caption{{Scatter plot of the specific angular momentum versus mass in the last epoch of the class I and earlier protostars. The red line is the least-squares fit ($\log y = (0.90\pm0.06) \, \log x + (20.8\pm0.04$)) to the data points, while the blue line is the fit to the simulated class II stars from Figure\,\ref{fig_j}. The gray dotted lines are the specific angular momentum for a given disk radius (1, 100, and 10000 au, with increasing $j$) as a function of stellar mass, using Eq.\,29 in \citet{Padoan+24_angular} for Keplerian disks.}}
    \label{fig_jb}
\end{figure}

This work stresses the importance of late BH accretion in class II disks. However, the same analysis can be applied to protostars. For that purpose, we select the stars that never reach the class II phase in the simulation, based on the criterion presented in \S\,\ref{sec:angular.1} and illustrated in Figure\,\ref{fig_time} (red dots in the figure). Similarly to Section\,\ref{sec:angular.2}, we compute $\boldsymbol j$ in the last ten epochs, and plot the value of $j$ of the last epoch versus the stellar mass in Figure\,\ref{fig_jb} for the embedded stars. The plot is similar to the case of class II stars. The main differences are the presence of more stars with masses around $10\,\rm M_{\odot}$ (most of the most massive stars are still forming at the end of the simulation), and a slightly shallower slope of the least-squares fit. The latter is primarily due to the presence of a significant number of low-mass stars with relatively large values of $j$ compared to the class II stars. Most of the low-mass class II stars have spent significant time in the BH phase, during which time their $v_{\rm rel}$ (velocity difference to the surrounding gas) has increased, as the stars have gradually decoupled from the surrounding gas. Increased $v_{\rm rel}$ reduces $R_{\rm BH}$ (see Eq.\,(\ref{R_BH})), hence decreasing the accreted $j$ \citep[see discussion of the time dependence in][]{Padoan+24_angular}. The low-mass class I stars, instead, being still fed by inertial inflows \citep{Padoan+20massive,Pelkonen+21}, and even more so in earlier protostellar phases, are more tightly coupled with the surrounding gas (smaller $v_{\rm rel}$), hence their $R_{\rm BH}$ is comparatively larger.

\begin{figure}
    \centering     
    \includegraphics[width=\hsize]{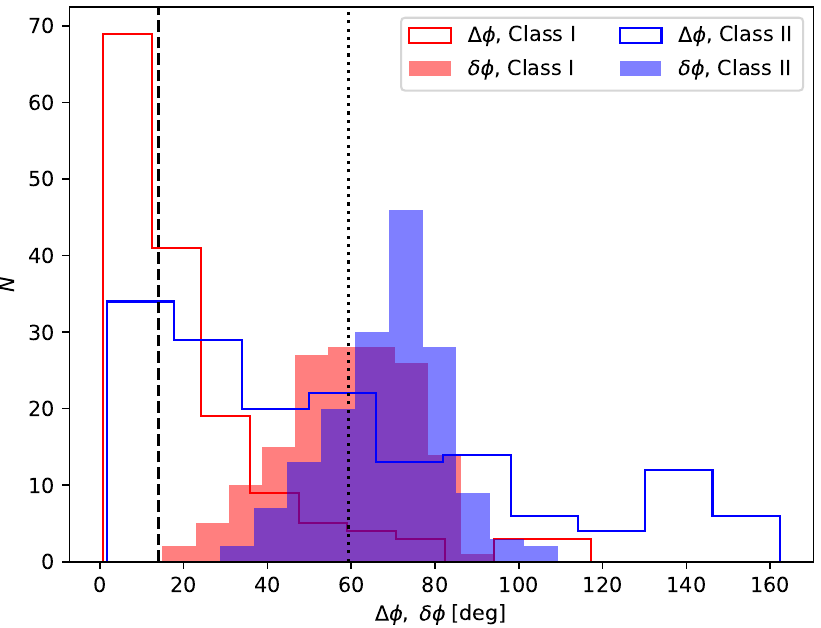}
    \caption{{Histograms of the last $\Delta \phi$ (red step histogram) and of the last $\delta \phi$ (red-shaded bar histogram), for class I and earlier protostars. The median values of $\Delta \phi$ and $\delta \phi$ are shown by the black dashed and dotted vertical lines, respectively. For comparison, the corresponding histograms for class II stars from Fig.~\ref{fig_angle_hist} have been replotted here in blue.}}
    \label{fig_angle_histb}
\end{figure}

A more striking difference between the accretion process during earlier protostellar phases (protostellar collapse and inertial inflows) and the class II phase (BH accretion) is illustrated in Figure\,\ref{fig_angle_histb}. The unshaded histograms compare the last $\Delta \phi$ distribution for the two classes, where $\Delta \phi$, as in Figure\,\ref{fig_angle_hist}, is the angle between the $\boldsymbol J$ vectors in the last two epochs. The histogram peaks at a much lower angle for the protostars compared to the class II stars, with median values of $14^{\circ}$ and $47^{\circ}$ respectively. This difference is to be expected, because the star-formation process is characterized by the protostellar collapse first, and by inertial inflows of spatially coherent dense filaments later, while BH accretion, after the star has decoupled from those converging flows (or said flows have ended), is a more stochastic process. As a result, disk misalignment is expected to be more likely in the later class II phase than during the star-formation phase. 

Figure\,\ref{fig_angle_histb} also shows the comparison between the $\delta \phi$ distributions, where $\delta \phi$ is a measure of the dispersion of the $\boldsymbol J$ vector orientation of individual tracer particles inside the last epoch. The difference between the two histograms is less pronounced, but in the same direction, as that between the $\Delta \phi$ histograms: class I stars have slightly smaller $\delta \phi$ values than class II stars, with median values of $59^{\circ}$ and $70^{\circ}$ respectively. The existence of a significant difference between the median values of $\delta \phi$ and $\Delta \phi$ also in the class I phase is expected, because the converging inertial flows forming the star come from different directions, each one bringing a different $\boldsymbol J$ orientation (as measured by individual tracer particles), even if they are rather coherent in time (hence the lower values of $\Delta \phi$).

\section{Synthetic observations}

We are proposing that late BH accretion plays a major role in the mass and angular momentum budget of class II disks. This scenario leads to disk infall rates on the order of the observed accretion rates of class II stars, and disk specific angular momentum and sizes consistent with the observations \citep[see also][]{Padoan+24_angular}. In this section we show that our main hypothesis can be directly tested observationally. We demonstrate that observations with the James Webb Space Telescope (JWST) can detect the predicted BH trails and also help to constrain their corresponding infall rates.

\subsection{Radiative transfer: SOC} 

We used the radiative transfer code SOC \citep{Juvela2019} to calculate images of NIR scattered light at 1.50\,$\mu$m and 2.77\,$\mu$m. The dust properties were based on the \citet{Weingartner2001} model with the selective extinction of $R_{\rm V}=5.5$. The observed level of scattered light depends mainly on the albedo and the extinction cross section. The albedo, $A$, of the $R_{\rm V}=5.5$ model is higher than in the corresponding model for normal Milky Way dust with $R_{\rm V}=3.1$ (e.g. 0.68 vs. 0.54 at 1.5\,$\mu$m). Such increase is expected inside molecular clouds as a result of grain evolution (ice deposition and grain coagulation), the effect increasing toward mid-infrared wavelengths \citep{Steinacker2010}. If the medium is optically thin at the wavelength of interest, a larger cross section (per gas mass) results in a higher intensity for the observed scattered light, but the effect would be the opposite for optically thick structures \citep{Juvela2020_L1642}. The NIR opacities of the $R_{\rm V}=3.1$ model would be some 30\% higher than in the adopted $R_{\rm V}=5.5$ model, where $\tau_{\rm ext}(1.5\,\mu{\rm m})=1$ is reached at $N({\rm H})=1.5\times 10^{22}\,{\rm cm}^{-2}$ and $\tau_{\rm ext}(2.77\,\mu{\rm m})=1$ at $N({\rm H})=3.8\times 10^{22}\,{\rm cm}^{-2}$. Typical streamers have lower column densities, and observed class II YSOs in Chamaeleon I \citep{Manara+17}, Upper Scorpius \citep{Manara+20}, and Taurus \citep{Gangi+22} have extinctions less than $A_V = 10 \rm \; mag$, typically from 1 to 5\,mag. Thus, although dust evolution may increase scattering via higher albedos, the negative effect of smaller absolute cross sections tends to be more significant.

To estimate the surface brightness due to the full line-of-sight, we started by calculating the scattered surface brightness for the full AMR model, using all AMR levels (smallest cell size of 25\,au). All stars were treated as separate point sources, and the model was further illuminated from the outside by the normal interstellar radiation field \citep{Mathis1983}. Maps of scattered light were registered separately for photons originating from the diffuse background and from the point sources. These showed that the surface brightness is smooth at the small scales relevant for the present study, apart from the signal produced by local embedded stars. 

For more detailed modeling of the environment around the stars (and to reach higher S/N for the resulting maps), we extracted from the full AMR grid smaller unigrid cubes that were each centered on a selected source (see below). The cubes had a size of 12800 au and a resolution of 50 au. The scattered light due to all embedded sources inside the cube was simulated using $8\times10^8$ photon packages. The numerical Monte Carlo noise in the resulting maps is mostly below 1\% of the mean surface brightness of the map, and the relative noise is lower for the detected streamers.

The luminosity and the temperature of the stars were calculated with the Modules for Experiments in Stellar Astrophysics (MESA) code \citep{Paxton+11,Paxton+13,Paxton+15,Paxton+18,Paxton+19,Jermyn+23}. The code has been augmented with modules for accretion of mass with a variable entropy and diagnostics to calculate the accretion luminosity, see \citet{Jensen+Haugboelle18} for details. In the following we will refer to the total stellar luminosity, the sum of photospheric and accretion luminosities, as the “luminosity” or “stellar luminosity."

\begin{figure}
\includegraphics[width=\hsize]{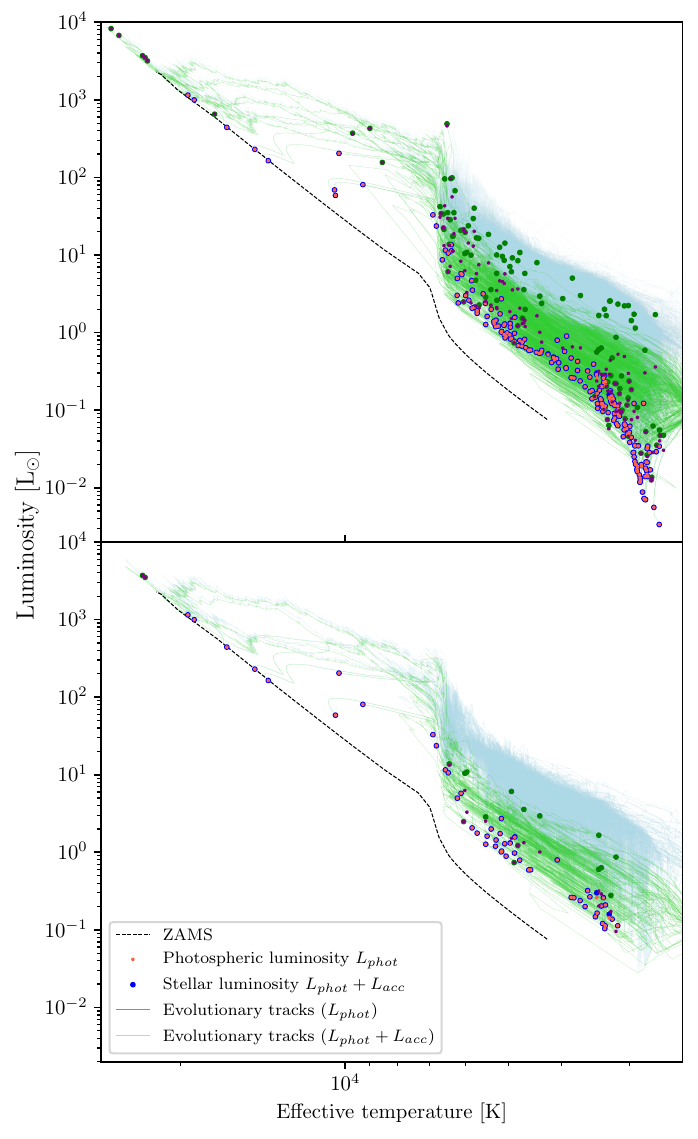}
\caption{{Hertzsprung-Russell diagram for all stars in the last snapshot (top panel) and for the 69 stars in the JWST synthetic observations (bottom panel). The small symbols (magenta and red) are for photospheric luminosities, while the large symbols (green and blue) are for stellar luminosities which include the accretion luminosity. Magenta and green points are class I stars, while red and blue points are class II. When $L_{\rm acc}$ is small (especially class II stars), the small points overlap with the larger ones, resulting in the appearance of red/magenta points with blue/green borders for class II and class I stars, respectively. The zero-age main sequence is indicated by the dashed black line, while the thin transparent light-green lines show the evolutionary tracks of all stars from an age of $10^4$ yr to the time of the snapshot, and the thin transparent light-blue lines are the evolutionary tracks with episodic accretion included. Stellar evolution is computed following the methodology of \citet{Jensen+Haugboelle18}.}}
\label{fig:HR}
\end{figure}

The Hertzsprung-Russell diagram for all stars in the last snapshot are shown in Fig.\ \ref{fig:HR}, where class I and class II sources are classified based on their local gas density: class I have local gas densities that are higher than $5\times 10^4$\,cm$^{-3}$, and class II have local gas densities that are lower (see Section\,\ref{sec:simulation}). At the time of the snapshot, only stars above 3.5 solar masses have reached the zero-age main sequence and commenced burning hydrogen, even while most of them are still deeply embedded. In light-green the evolutionary tracks, including self-consistent account for mass and entropy accretion, highlighting the protostellar evolution. For 23\% (15\%) of the stars the accretion luminosity, $L_{\rm acc}$, contribute more than 20\% (50\%) of the total luminosity, as can be seen by considering offsets between the magenta and green points for class I stars (the red and blue points for class II) in Fig.\ \ref{fig:HR}.

The stars of interest were selected from the last snapshot of the simulation, with a luminosity cut ($L > 0.1 \, \rm L_\odot$) and an accretion-rate cut ($10^{-11} < \dot{M} < 10^{-6}  \, \rm M_\odot/yr$), irrespective of their classification as class I or class II stars. Based on an initial set of calculations, stars less luminous than this are not able to illuminate their surroundings sufficiently, and those with lower accretion rates do not have enough dust to scatter their light. All the other stars within the computational volume were included regardless of their luminosity and accretion rate, as they may still contribute to the surface brightness of the scattered light. However, this also means that we need to be careful about attributing a detection when it is due to a nearby star. To alleviate this concern, in the cases where we have more than one star in the subcube, we consider only the brightest one (called the dominant star). This results in 48 single stars and 21 dominant stars, observed from three directions with two wavelengths, for a total of 207 maps (69 stars times three directions) for each of the two wavelengths and for the column density.

\subsection{Scattered NIR light with JWST}

\begin{figure}
\includegraphics[width=\hsize]{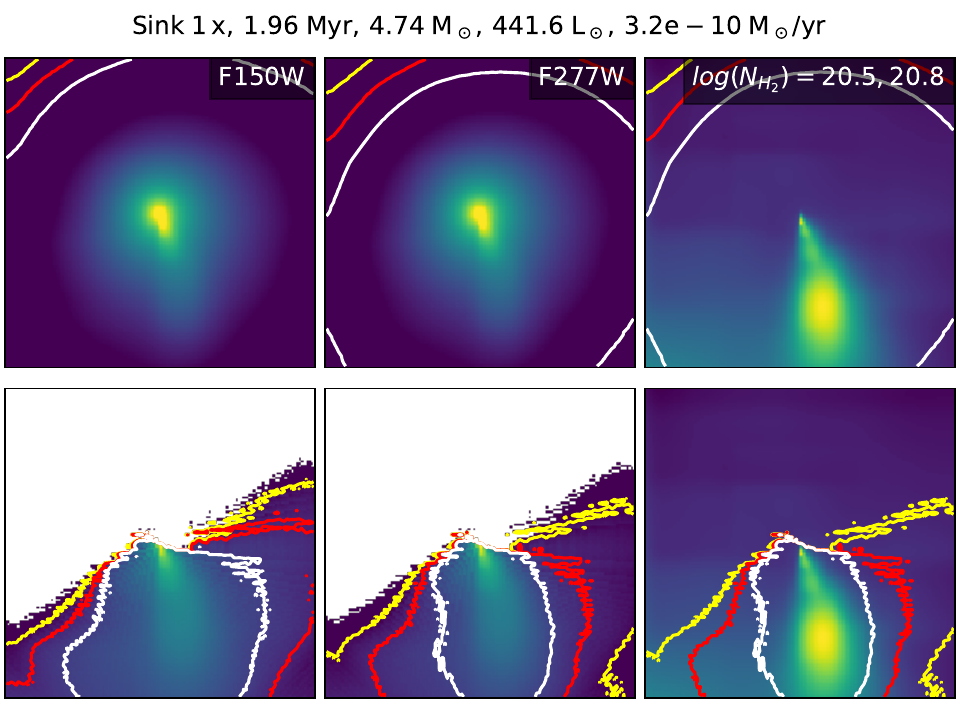}
\caption{{JWST synthetic NIR observations (see Section 4) around a relatively high-mass star (located at the center of the map) in a 6400 x 6400\,au frame. The title lists the sink particle number and direction of projection, the age, the mass, the luminosity and the accretion rate. The top row shows, in log scale the surface brightness in the two selected wavelengths (left and central images respectively, with the JWST filter as a label), and and the local column density integrated across the 12800\,au depth of the selected volume (right image, with minimum and maximum log value of the color scale as a label). The contours (white, red, and yellow) are $5\sigma$ limits for 2, 8 and 50 min exposure time, respectively (the column density image shows the F277W contours). The bottom row shows the corresponding surface brightness images after the ring subtraction described in the text (the column density map is the same as in the upper row). Negative and zero pixel values are masked out (white background). Contours show the 5-sigma limits of the remaining surface brightness, after the ring subtraction.}}
\label{fig:sink_1}
\end{figure} 

\begin{figure}
\includegraphics[width=\hsize]{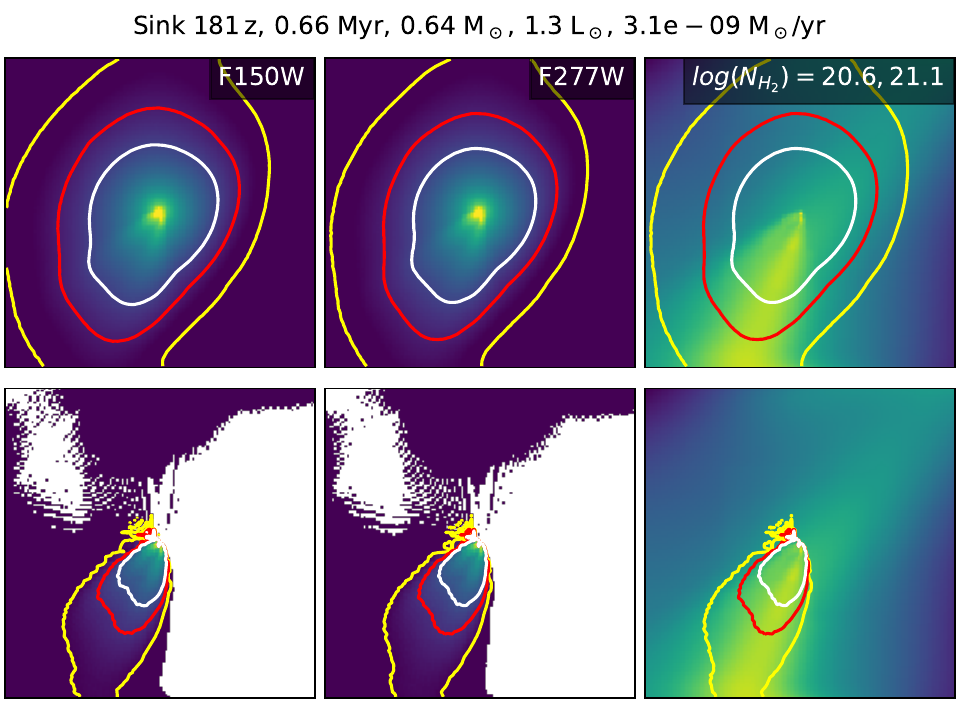}
\caption{{As Fig.~\ref{fig:sink_1}, but for a lower-mass star.}}
\label{fig:sink_181}
\end{figure} 

The 414 surface brightness maps are included with the corresponding 207 column density maps for the analysis. We assume that the distance to the target cloud is 200\,pc, close to the distance of Chamaeleon I. We then perform a very simple ring subtraction to get rid of symmetrical features: we divide the map into 1-pixel thick rings around the star and subtract the median of the surface brightness within each of the rings from the pixels in the corresponding ring.
This highlights the asymmetric structures that are expected from BH accretion in class II stars, or from accreting filaments in class I stars. As a very simple way to analyze the success of the detection, we calculate the signal-to-noise ratio (S/N) using the Exposure Time Calculator of James Webb Space Telescope (JWST). We assume a rapid readout, leading to an exposure time of 107.4 seconds (about two minutes), and an aperture size of 0.125" to match with our map pixel size, which results in 1$\sigma$ noise levels of 0.043 MJy/sr and 0.0074 MJy/sr for filters F150W and F277W, respectively. We also test longer exposure times of about 8 minutes and 50 minutes. If we have traced a continuous structure from the center to at least 1000 au at S/N $\ge 5\sigma$, then we consider it a successful detection of an asymmetric structure, hence a candidate for an accretion flow. In addition, we measure the surface brightness within a ring between 250 au and 1000 au, in order to compare it with the accretion rate and luminosity of the star.

As the MHD model did not have disks, we are unable to model disk shadows. We also did not consider the effect of the occulting masks for coronagraphy in order to block the central star, which will also cause some limits on the usable filters. However, as we are looking at larger distances from the central star, the results are still valid.

\begin{figure}
\includegraphics[width=\hsize]{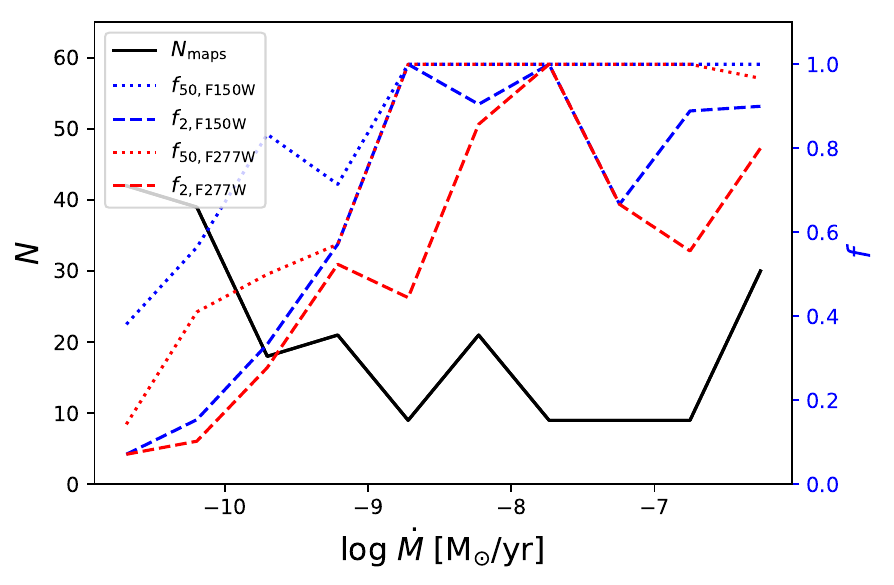}
\caption{{Total number of maps in bins of the accretion rate, and the detection fractions of the maps where accretion features were detected with S/N $\ge 5\sigma$ at least 1000\,au from the central star. The black line is the number of all the maps, non-detections included, with y-axis on the left. The colored lines (blue and red for F150W and F277W filters, and dashed and dotted for 2 min and 50 min exposure times, respectively) and the y-axis on the right show the successful detection fraction of the maps.}}
\label{acc_stream}
\end{figure} 

\begin{figure}
\includegraphics[width=\hsize]{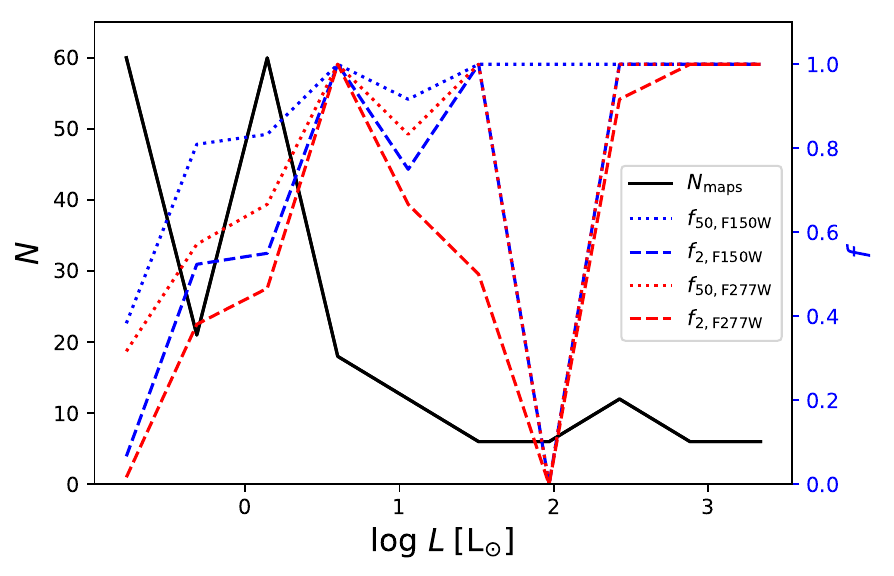}
\caption{{As Fig.~\ref{acc_stream}, but in bins of the stellar luminosity instead of the accretion rate.}}
\label{luminosity_stream}
\end{figure}

Figure~\ref{fig:sink_1} shows a relatively high-mass star ($4.7 \, \rm M_\odot$), accreting at a rate of $3.2 \times 10^{-10} \, \rm M_\odot / yr$. Despite its modest accretion rate, its high luminosity illuminates the surrounding dust, but also the streamer seen in the column density map, as the NIR photons scatter from the dust particles. The resulting surface brightness is enough that even after we perform a very rudimentary ring subtraction, the remaining surface brightness structure aligns almost perfectly with the BH trails seen in the column density map, leading us to conclude that this is a successful detection. The white contour shows that even a two-minutes exposure with JWST would be enough for this detection.

Figure~\ref{fig:sink_181} shows a lower-mass star ($0.64 \, \rm M_\odot$), accreting at a rate of $3.1 \times 10^{-9} \, \rm M_\odot / yr$. Its luminosity is only slightly higher than that of the Sun, but it is enough for the BH accretion feature to be detected. The opening angle of the BH wake is larger than in the previous case, indicating a smaller relative velocity between the star and the gas.

\begin{figure*}
\includegraphics[width=17cm]{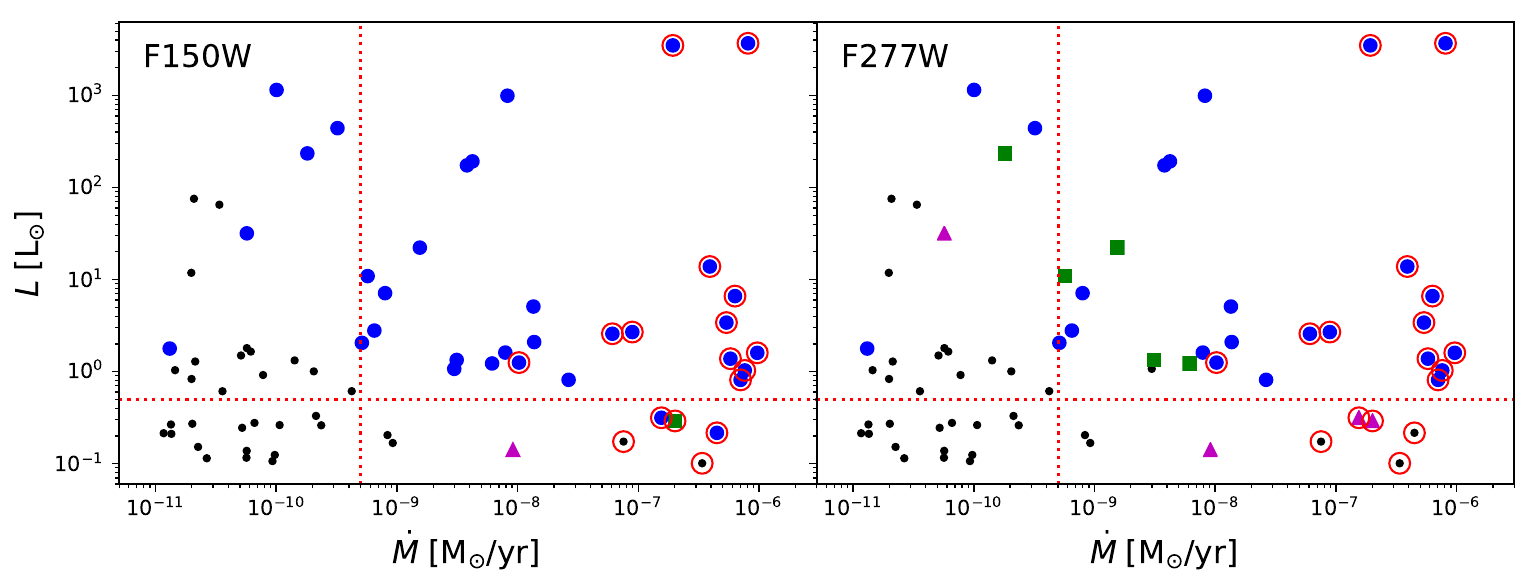}
\caption{{Scatter plots of stellar luminosity versus accretion rate for all the synthetic observations, with symbols and colors referring to the detection rate after 2\,min of exposure time with the F150W (left panel) and F277W (right panel) filters. The small black dots correspond to a non-detection of an accretion feature in all three directions ($x$, $y$, and $z$), the magenta triangles detection in only one direction, the green squares detection in two directions, and the large blue dots detection in all three directions. The protostars (class I or earlier) are highlighted by red circles around the markers. The red dotted vertical and horizontal lines mark the thresholds of $\dot{M} > 5 \times 10^{-10}  \, \rm M_\odot/yr$ and $L > 0.5 \, \rm L_\odot$, respectively.}}
\label{acc_L_2min}
\end{figure*} 

Figures~\ref{acc_stream} and~\ref{luminosity_stream} show the total number of maps in bins of the accretion rate and the luminosity of the sources, respectively, and the fraction of the successful detections of accretion features for both filters, using 2 min and 50 min exposure times. Obviously, the detection fraction is higher for longer exposure time and higher luminosity. It is also higher for the shorter wavelength, due to the increased stellar flux, and higher accretion rate, because that usually corresponds to higher column density in the accreting gas. The drop in the detection fraction at $100 \, \rm L_\odot$ is due to two sources with very low accretion rates, as shown in Figure~\ref{acc_L_2min}. Even with just 2\,min exposure, JWST's detection fraction rises above 50\% for $\dot{M}\ge 10^{-9} \, \rm M_\odot/yr$, or for $L\ge 1 \, \rm L_\odot$. The longer exposures can push the 50\% success rate to smaller values of $\dot{M}$ and  $L$. Reducing the resolution of the map by averaging over several pixels would improve the signal-to-noise as well, at the cost of the resolution, as long as the resolution is still high enough to resolve the asymmetric structures.

Figure~\ref{acc_L_2min} shows scatter plots of the accretion rate and luminosity for all sources. The colors and the symbols refer to the detection rate per source with a 2\,min exposure time and using either F150W (left panel) or F277W filter (right panel): a small black dot corresponds to a non-detection in all three maps ($x$, $y$, and $z$ directions), a magenta triangle to a detection in one map out of three, a green square to a detection in two maps out of three, and a large blue dot to a detection in all three maps. The protostars (class I and earlier) are highlighted with red circles around the markers. The figure shows that a selection based on both accretion rate and luminosity thresholds improves the detection fraction. If we require both $\dot{M}> 5 \times 10^{-10}  \, \rm M_\odot/yr$ and $L > 0.5 \, \rm L_\odot$ (red dotted lines), the detection fraction is 100\% for F150W, regardless of further cuts by protostellar classification or maximum luminosity. For F277W filter with the same requirements, the detection fraction is approximately 90\%, 74 detections out of 81 maps. If we consider only the class II stars, the detection fraction is still 84\%, 38 detections out of 45 maps. If we also discard the three class II stars with $L > 100 \, \rm L_\odot$, to reflect the scarcity of such sources in the solar neighborhood, we get a detection fraction of  81\%, 29 detections out of 36 maps. 

\begin{figure*}
\includegraphics[width=17cm]{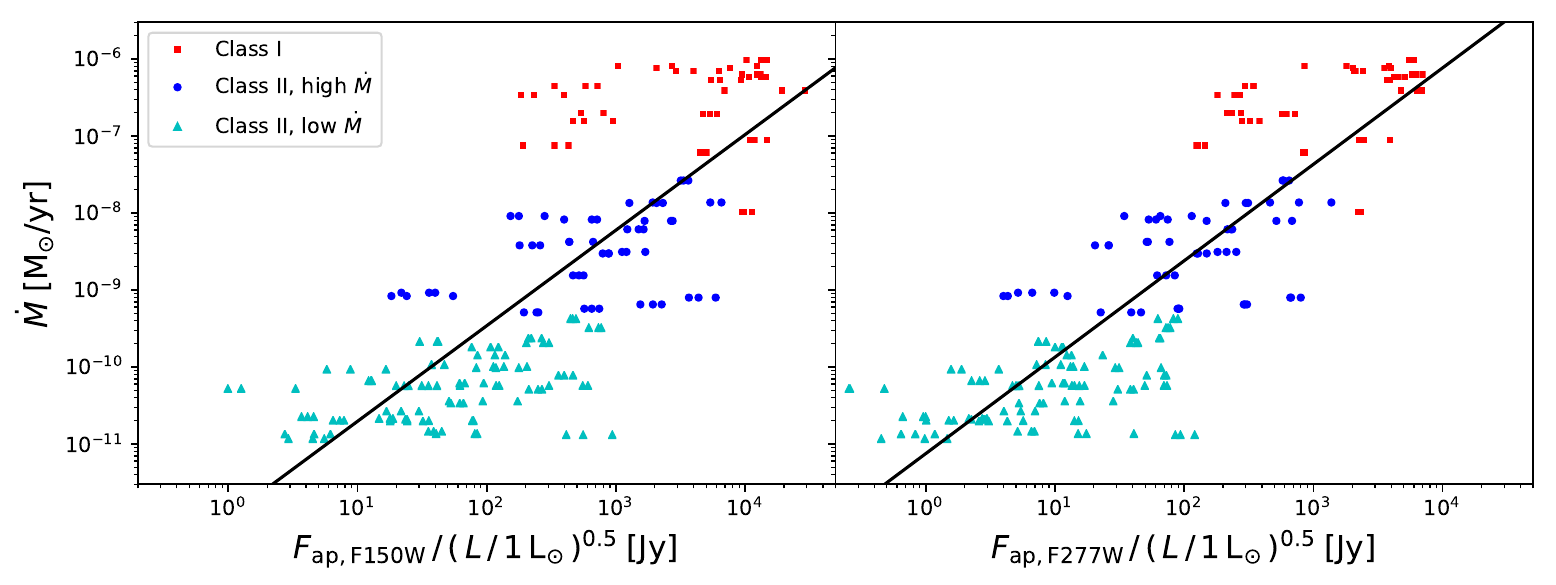}
\caption{{Comparison of the NIR flux density (F150W (left panel) or F277W (right panel) filter, aperture from 250 to 1000\,au), divided by the square root of the luminosity of the star in solar luminosities, to the accretion rate. Blue dots ($\dot{M} \ge 5 \times 10^{-10} \; \rm M_\odot/yr$) and cyan triangles ($\dot{M} < 5 \times 10^{-10} \; \rm M_\odot/yr$) are for the class II stars (local density $\le 5\times 10^4$\,cm$^{-3}$), and the small red squares are for the class I stars (local density $> 5\times 10^4$\,cm$^{-3}$). The black solid line is a least-squares fit giving $\log y = (1.24\pm0.07) \, \log x - (11.9\pm0.18)$ (F150W, left panel) and $\log y = (1.25\pm0.05) \, \log x - (11.1\pm0.11)$ (F277W, right panel).}}
\label{F277W_compare}
\end{figure*} 

In Figure~\ref{F277W_compare} we compare the NIR flux density in an aperture from 250 to 1000 au with both filters to the accretion rate. We do not find a significant difference between the distribution of single and dominant stars, so we do not differentiate between them in the plot. Instead, we distinguish between class I and class II, and also split the class II stars into high and low accretion rates, based on the criterion of $\dot{M} > 5 \times 10^{-10}  \, \rm M_\odot/yr$ found in the previous paragraph. We find that there is a large scatter in flux density between stars of a similar accretion rate, which is partially explained by the luminosity differences between those stars. Thus, we divide the flux density with the square root of the luminosity, which helps to reduce the scatter and separate the different accretion regimes, especially when using F277W. Figure~\ref{F277W_compare} shows a clear relationship between the accretion rate and the measured flux density, which is to be expected, as the higher accretion rate implies a denser medium that the star is traveling through and a denser accretion flow, meaning that there is more dust to scatter the NIR photons. We fit a power-law to each panel, which could be used as a rough order-of-magnitude estimate of the accretion rate using the two observables, the stellar luminosity and the F277W flux density, even if the surface brightness of the BH wake itself were too weak for a high signal-to-noise detection. However, more complicated radiative transfer modeling would be needed to accurately derive the accretion rate from the surface brightness, possibly using the morphology of the surface brightness structure rather than simple aperture photometry.

\section{Discussion}

This work provides further evidence that late BH accretion on class II stars can fully replenish their disks. More importantly, thanks to its Lagrangian analysis with tracer particles, it shows that the random nature of BH accretion results in significant time variations of the angular momentum direction of the accreted gas. Such variations may cause angular momentum cancelation in the disk, instabilities and pressure traps, affecting the disk evolution and its accretion rate onto the star. They may also result in observable evidence of disk misalignment, such as misalignment of inner and outer disks \citep{Marino+15,Pinilla+18,Ansdell+20}, misaligned single and multi-planet systems \citep{Hubert+13,Sanchis-Ojeda+13,Kamiaka+19,Hjorth+21,Albrecht+22}, distributions of projected stellar rotation velocities of stars with transiting planets \citep{Louden+21,Esteves+23}, hot Jupiters with orbit obliquity beyond the Kozai angle \citep{Kawai+24}, and possibly misalignment between the disk and the stellar spin in class II stars \citep[too high disk extinction of class II stars with rotation-axis inclination larger than $\sim 60^{\circ}$,][]{Kounkel+23}. Evidence for the warped or misaligned disks \citep{Sai+20,Yamato+23} and misaligned outflows \citep{Okoda+21,Sai+24} have also been found in earlier star formation phases.

This scenario of late-time mass infall is consistent with recent discoveries of large-scale flows (streamers) around class II disks \citep{Ginski+21,Huang+21,Valdivia+24}, as well as the general detection of reflection nebulae around class II stars \citep{Gupta+23}. This suggests that targeted observations, such as the JWST observations discussed here, may result in a high detection rate of streamers around class II stars as well. Follow-up ALMA observations may then probe the kinematics of class II streamers, to be compared with accreting filaments in BH trails seen in the simulations.
Such large-scale flows have also been found around class 0 stars \citep[e.g.,][]{Thieme+22,Kido+2023,Aso+2023}, and around class I stars or stars in transition between class I and class II \citep[e.g.,][]{Yen+19,Pineda+20,Alves+20,Grant+21,Valdivia+22,Cacciapuoti+24,Valdivia+24}.

We have derived a dependence of $j$ on the stellar mass, as shown in Figure\,\ref{fig_j} and Eq.\,(\ref{eq_j_M}). As commented in \S\,\ref{sec:angular.2}, the derived values of $j$ are close to those from our Eulerian analysis (Padoan et al. 2024) and only a factor of two larger than those inferred from the observed disk sizes, which is to be expected because $j$ is most likely not perfectly conserved as material settles to the disk from the BH radius (a factor of four contraction -- see Eq.\,(\ref{R_d_BH_t_main})). A correlation between observed disk sizes and stellar mass, for class II stars, qualitatively consistent with our result, was first suggested by \citet{Andrews+2018}, and later confirmed in other works \citep[e.g.,][]{Hendler+20,Andrews20,Long+22}. The reported slope of the dependence of disk size on stellar mass is different in different observational samples, varying from 0.3 in \citet{Long+22} to 0.9 in \citet{Andrews20}. In our sample that combines results from different surveys we find a slope of $\sim 0.4\pm0.1$. 

Because $j\sim M_{\rm star}^{1/2} \, R_{\rm d}^{1/2}$ in a Keplerian disk, our model prediction of Eq.\,(\ref{j_BH_t_main}), $j\sim M_{\rm star}$, corresponds to $R_{\rm d}\sim M_{\rm star}$. The result of the Lagrangian analysis of this work corresponds to a slightly steeper relation, with an exponent of 1.28, while the Eulerian analysis in Padoan et al. (2024) results in a slightly shallower exponent of 0.72. At face value, the model prediction is steeper than the observed relation (though comparable in the case of \citet{Andrews20}). However, the observed relation is affected by a significant observational bias: the fraction of unresolved disks with sizes smaller than $\sim 20$\,au in the surveys is very large, $\sim 70$\% according to \citet{Manara+23}, and clearly skewed toward lower stellar masses (expected to host the smaller disk sizes), so the real mass dependence of the disk sizes should be steeper than what is currently derived from the observations.  

We note that a relation between disk size and stellar mass has been recently found also from observations of class 0 and I disks, with a reported slope of 1.1 in \citet{Yen+24}, and 0.7 in \citet{YenLee24}. Our fit for embedded stars in Figure\,\ref{fig_jb} corresponds to the relation $R_{\rm d}\sim M_{\rm star}^{0.8}$, consistent with the observations. However, we stress that we have not attempted to predict the value of $j$ for actual class 0 and I disks, because $j$ (or the size) of very young disks must have a significant contribution from the initial collapse of the prestellar core, so it should be primarily determined not by the mass infall at a specific time, which is what we compute here, but by the cumulative infall up to that time. The disk size is in fact expected to grow during the collapse of a prestellar core, though with a shallower mass dependence than observed (an exponent $\le 0.5$), depending on the initial orientation of the magnetic field \citep[e.g.,][]{Hennebelle+16,Lee+21}. For class II stars, over Myr timescales the Bondi-Hoyle infall (from the ISM to the disks), comparable to the observed accretion rates (from the disks to the stars) as shown in Fig.\,\ref{fig_accretion}, can fully rejuvenate the disks, so the disk properties inherited from the initial prestellar-core collapse are less relevant.

We find that the angular momentum is not conserved from larger to smaller scales. We should in general expect magnetic breaking to play a role in this MHD simulation, as well as in nature, causing a transfer of angular momentum to larger scales (see for example the MHD disk-formation models by \citet{Hennebelle+16} and \citet{Lee+21}). However, observations of L1527 IRS, a class 0 protostar, by \citet{Ohashi+14}, and of L1489 IRS, a class I protostar, by \citet{Sai+20} suggest that the specific angular momentum of infalling materials may be conserved within infalling envelopes ($\sim 1000$\,au scale). Even if we assumed that angular momentum were conserved from the Bondi-Hoyle radius to the disk radius, this would only scale up the specific angular momenta found in this study by a factor of two, meaning that there would be an even larger angular momentum budget from late-stage infall to explain the observed class II disk sizes.

\section{Conclusions}

Using a star-formation simulation with a spatial resolution of 25\,au, where the full stellar IMF is successfully reproduced, but stellar disks are not resolved, we have studied the angular momentum of the gas gravitationally captured by the stars, with a Lagrangian analysis based on tracer particles. Our main results are summarized in the following.

\begin{enumerate}

\item BH infall rates of class II stars in the simulation are of the same order as the observed accretion rates of class II stars, suggesting that class II disks may be replenished from larger scale at a similar mass rate as they accrete onto the central star. 

\item The specific angular momentum of the gas gravitationally captured by class II stars scales approximately linearly with the stellar mass, in agreement with the analytical prediction and numerical analysis in \citet{Padoan+24_angular}, and with the observational $j-M$ relation for spatially resolved disks.    

\item In class II stars, the direction of the accreted angular momentum may vary strongly with time, with a median angle of $47^{\circ}$ between consecutive epochs where the accreted mass is of the order of one disk mass. The time evolution is more coherent in the class I phase, with a median angle of $14^{\circ}$.
    
\item Synthetic observations of class II stars in the simulation with accretion rates $\dot{M}> 5 \times 10^{-10}  \, \rm M_\odot/yr$ show that their BH accretion flows can be detected by JWST in NIR scattered light. Imposing an additional selection constraint for the stellar luminosity of $L > 0.5 \, \rm L_\odot$, the estimated detection success rate with JWST is over 80\% with both filters (100\% in F150W) and only 2 minutes of exposure time.

\end{enumerate}

With regard to the mass budget, BH accretion in class II stars may alleviate the problems of the short disk accretion times relative to the disk ages \citep[][]{Fedele+10,Manara+20,Mauco+23}, and of the insufficient disk masses relative to the masses of exoplanetary systems \citep[e.g.,][]{Ansdell+2016,Manara+18,Mulders+21,Stefansson+23}. As for the angular momentum evolution, the contribution of BH accretion in the class II phase cannot be neglected, as it provides a large enough angular momentum to explain the observed size of class II disks, as well as significant time variations to induce strong perturbations in the disks.   

Our scenario for class II disks leads to the prediction that, for a high enough accretion rate and stellar luminosity, accretion streamers should be readily detected by targeted JWST observations. If confirmed by future observations, this scenario would have important implications for models of planet formation and disk evolution.

\begin{acknowledgements}
    We thank the anonymous referee for their helpful comments, leading to an improved and more clear presentation of our work. VMP and PP acknowledge financial support by the grant PID2020-115892GB-I00, funded by MCIN/AEI/10.13039/501100011033 and by the grant CEX2019-000918-M funded by MCIN/AEI/10.13039/501100011033. VMP gratefully acknowledges also financial support from the European Research Council via the ERC Synergy Grant "ECOGAL" (project ID 855130). MJ acknowledges support from the Research Council of Finland grant No. 348342. The research leading to these results has received funding from the Independent Research Fund Denmark through grant No. DFF 802100350B (TH). We acknowledge PRACE for awarding us access to Joliot-Curie at GENCI@CEA, France. The astrophysics HPC facility at the University of Copenhagen, supported by research grants from the Carlsberg, Novo, and Villum foundations, was used for carrying out the postprocessing, analysis, and long-term storage of the results.
\end{acknowledgements}


\bibliographystyle{aa}
\bibliography{aanda}

\begin{thebibliography}{91}
\expandafter\ifx\csname natexlab\endcsname\relax\def\natexlab#1{#1}\fi

\bibitem[{{Albrecht} {et~al.}(2022){Albrecht}, {Dawson}, \& {Winn}}]{Albrecht+22}
{Albrecht}, S.~H., {Dawson}, R.~I., \& {Winn}, J.~N. 2022, \pasp, 134, 082001

\bibitem[{{Alves} {et~al.}(2020){Alves}, {Cleeves}, {Girart}, {Zhu}, {Franco}, {Zurlo}, \& {Caselli}}]{Alves+20}
{Alves}, F.~O., {Cleeves}, L.~I., {Girart}, J.~M., {et~al.} 2020, \apjl, 904, L6

\bibitem[{{Andrews}(2020)}]{Andrews20}
{Andrews}, S.~M. 2020, \araa, 58, 483

\bibitem[{{Andrews} {et~al.}(2018){Andrews}, {Huang}, {P{\'e}rez}, {Isella}, {Dullemond}, {Kurtovic}, {Guzm{\'a}n}, {Carpenter}, {Wilner}, {Zhang}, {Zhu}, {Birnstiel}, {Bai}, {Benisty}, {Hughes}, {{\"O}berg}, \& {Ricci}}]{Andrews+2018}
{Andrews}, S.~M., {Huang}, J., {P{\'e}rez}, L.~M., {et~al.} 2018, \apjl, 869, L41

\bibitem[{{Ansdell} {et~al.}(2020){Ansdell}, {Gaidos}, {Hedges}, {Tazzari}, {Kraus}, {Wyatt}, {Kennedy}, {Williams}, {Mann}, {Angelo}, {D{\^u}chene}, {Mamajek}, {Carpenter}, {Esplin}, \& {Rizzuto}}]{Ansdell+20}
{Ansdell}, M., {Gaidos}, E., {Hedges}, C., {et~al.} 2020, \mnras, 492, 572

\bibitem[{{Ansdell} {et~al.}(2016){Ansdell}, {Williams}, {van der Marel}, {Carpenter}, {Guidi}, {Hogerheijde}, {Mathews}, {Manara}, {Miotello}, {Natta}, {Oliveira}, {Tazzari}, {Testi}, {van Dishoeck}, \& {van Terwisga}}]{Ansdell+2016}
{Ansdell}, M., {Williams}, J.~P., {van der Marel}, N., {et~al.} 2016, \apj, 828, 46

\bibitem[{{Aso} {et~al.}(2023){Aso}, {Kwon}, {Ohashi}, {J{\o}rgensen}, {Tobin}, {Aikawa}, {de Gregorio-Monsalvo}, {Han}, {Kido}, {Koch}, {Lai}, {Lee}, {Lee}, {Li}, {Lin}, {Looney}, {Narayanan}, {Phuong}, {Sai}, {Saigo}, {Santamar{\'\i}a-Miranda}, {Sharma}, {Takakuwa}, {Thieme}, {Tomida}, {Williams}, \& {Yen}}]{Aso+2023}
{Aso}, Y., {Kwon}, W., {Ohashi}, N., {et~al.} 2023, \apj, 954, 101

\bibitem[{{Baraffe} {et~al.}(2015){Baraffe}, {Homeier}, {Allard}, \& {Chabrier}}]{Baraffe+15}
{Baraffe}, I., {Homeier}, D., {Allard}, F., \& {Chabrier}, G. 2015, \aap, 577, A42

\bibitem[{{Bondi}(1952)}]{Bondi52}
{Bondi}, H. 1952, \mnras, 112, 195

\bibitem[{{Cacciapuoti} {et~al.}(2024){Cacciapuoti}, {Macias}, {Gupta}, {Testi}, {Miotello}, {Espaillat}, {K{\"u}ffmeier}, {van Terwisga}, {Tobin}, {Grant}, {Manara}, {Segura-Cox}, {Wendeborn}, {Klessen}, {Maury}, {Lebreuilly}, {Hennebelle}, \& {Molinari}}]{Cacciapuoti+24}
{Cacciapuoti}, L., {Macias}, E., {Gupta}, A., {et~al.} 2024, \aap, 682, A61

\bibitem[{{Chabrier}(2005)}]{Chabrier05}
{Chabrier}, G. 2005, in Astrophysics and Space Science Library, Vol. 327, The Initial Mass Function 50 Years Later, ed. E.~{Corbelli}, F.~{Palla}, \& H.~{Zinnecker}, 41

\bibitem[{{Cieza} {et~al.}(2021){Cieza}, {Gonz{\'a}lez-Ruilova}, {Hales}, {Pinilla}, {Ru{\'\i}z-Rodr{\'\i}guez}, {Zurlo}, {Casassus}, {P{\'e}rez}, {C{\'a}novas}, {Arce-Tord}, {Flock}, {Kurtovic}, {Marino}, {Nogueira}, {Perez}, {Price}, {Principe}, \& {Williams}}]{Cieza+21}
{Cieza}, L.~A., {Gonz{\'a}lez-Ruilova}, C., {Hales}, A.~S., {et~al.} 2021, \mnras, 501, 2934

\bibitem[{{Esteves} {et~al.}(2023){Esteves}, {Izidoro}, {Winter}, {Bitsch}, \& {Isella}}]{Esteves+23}
{Esteves}, L., {Izidoro}, A., {Winter}, O.~C., {Bitsch}, B., \& {Isella}, A. 2023, \mnras, 521, 5776

\bibitem[{{Evans} {et~al.}(2009){Evans}, {Dunham}, {J{\o}rgensen}, {Enoch}, {Mer{\'{\i}}n}, {van Dishoeck}, {Alcal{\'a}}, {Myers}, {Stapelfeldt}, {Huard}, {Allen}, {Harvey}, {van Kempen}, {Blake}, {Koerner}, {Mundy}, {Padgett}, \& {Sargent}}]{Evans+09}
{Evans}, II, N.~J., {Dunham}, M.~M., {J{\o}rgensen}, J.~K., {et~al.} 2009, \apjs, 181, 321

\bibitem[{{Fedele} {et~al.}(2010){Fedele}, {van den Ancker}, {Henning}, {Jayawardhana}, \& {Oliveira}}]{Fedele+10}
{Fedele}, D., {van den Ancker}, M.~E., {Henning}, T., {Jayawardhana}, R., \& {Oliveira}, J.~M. 2010, \aap, 510, A72

\bibitem[{{Feiden}(2016)}]{Feiden+16}
{Feiden}, G.~A. 2016, \aap, 593, A99

\bibitem[{{Gangi} {et~al.}(2022){Gangi}, {Antoniucci}, {Biazzo}, {Frasca}, {Nisini}, {Alcal{\'a}}, {Giannini}, {Manara}, {Giunta}, {Harutyunyan}, {Munari}, \& {Vitali}}]{Gangi+22}
{Gangi}, M., {Antoniucci}, S., {Biazzo}, K., {et~al.} 2022, \aap, 667, A124

\bibitem[{{Ginski} {et~al.}(2021){Ginski}, {Facchini}, {Huang}, {Benisty}, {Vaendel}, {Stapper}, {Dominik}, {Bae}, {M{\'e}nard}, {Muro-Arena}, {Hogerheijde}, {McClure}, {van Holstein}, {Birnstiel}, {Boehler}, {Bohn}, {Flock}, {Mamajek}, {Manara}, {Pinilla}, {Pinte}, \& {Ribas}}]{Ginski+21}
{Ginski}, C., {Facchini}, S., {Huang}, J., {et~al.} 2021, \apjl, 908, L25

\bibitem[{{Grant} {et~al.}(2021){Grant}, {Espaillat}, {Wendeborn}, {Tobin}, {Mac{\'\i}as}, {Rilinger}, {Ribas}, {Megeath}, {Fischer}, {Calvet}, \& {Hee Kim}}]{Grant+21}
{Grant}, S.~L., {Espaillat}, C.~C., {Wendeborn}, J., {et~al.} 2021, \apj, 913, 123

\bibitem[{{Gupta} {et~al.}(2023){Gupta}, {Miotello}, {Manara}, {Williams}, {Facchini}, {Beccari}, {Birnstiel}, {Ginski}, {Hacar}, {K{\"u}ffmeier}, {Testi}, {Tychoniec}, \& {Yen}}]{Gupta+23}
{Gupta}, A., {Miotello}, A., {Manara}, C.~F., {et~al.} 2023, \aap, 670, L8

\bibitem[{{Haugb{\o}lle} {et~al.}(2018){Haugb{\o}lle}, {Padoan}, \& {Nordlund}}]{Haugbolle+18imf}
{Haugb{\o}lle}, T., {Padoan}, P., \& {Nordlund}, {\r{A}}. 2018, \apj, 854, 35

\bibitem[{{Hendler} {et~al.}(2020){Hendler}, {Pascucci}, {Pinilla}, {Tazzari}, {Carpenter}, {Malhotra}, \& {Testi}}]{Hendler+20}
{Hendler}, N., {Pascucci}, I., {Pinilla}, P., {et~al.} 2020, \apj, 895, 126

\bibitem[{{Hennebelle} {et~al.}(2016){Hennebelle}, {Commer{\c{c}}on}, {Chabrier}, \& {Marchand}}]{Hennebelle+16}
{Hennebelle}, P., {Commer{\c{c}}on}, B., {Chabrier}, G., \& {Marchand}, P. 2016, \apjl, 830, L8

\bibitem[{{Heyer} \& {Brunt}(2004)}]{Heyer+Brunt04}
{Heyer}, M.~H. \& {Brunt}, C.~M. 2004, \apjl, 615, L45

\bibitem[{{Hjorth} {et~al.}(2021){Hjorth}, {Albrecht}, {Hirano}, {Winn}, {Dawson}, {Zanazzi}, {Knudstrup}, \& {Sato}}]{Hjorth+21}
{Hjorth}, M., {Albrecht}, S., {Hirano}, T., {et~al.} 2021, Proceedings of the National Academy of Science, 118, e2017418118

\bibitem[{{Hoyle} \& {Lyttleton}(1939)}]{Hoyle+Lyttleton39}
{Hoyle}, F. \& {Lyttleton}, R.~A. 1939, Proceedings of the Cambridge Philosophical Society, 35, 405

\bibitem[{{Huang} {et~al.}(2021){Huang}, {Bergin}, {{\"O}berg}, {Andrews}, {Teague}, {Law}, {Kalas}, {Aikawa}, {Bae}, {Bergner}, {Booth}, {Bosman}, {Calahan}, {Cataldi}, {Cleeves}, {Czekala}, {Ilee}, {Le Gal}, {Guzm{\'a}n}, {Long}, {Loomis}, {M{\'e}nard}, {Nomura}, {Qi}, {Schwarz}, {Tsukagoshi}, {van't Hoff}, {Walsh}, {Wilner}, {Yamato}, \& {Zhang}}]{Huang+21}
{Huang}, J., {Bergin}, E.~A., {{\"O}berg}, K.~I., {et~al.} 2021, \apjs, 257, 19

\bibitem[{{Huber} {et~al.}(2013){Huber}, {Carter}, {Barbieri}, {Miglio}, {Deck}, {Fabrycky}, {Montet}, {Buchhave}, {Chaplin}, {Hekker}, {Montalb{\'a}n}, {Sanchis-Ojeda}, {Basu}, {Bedding}, {Campante}, {Christensen-Dalsgaard}, {Elsworth}, {Stello}, {Arentoft}, {Ford}, {Gilliland}, {Handberg}, {Howard}, {Isaacson}, {Johnson}, {Karoff}, {Kawaler}, {Kjeldsen}, {Latham}, {Lund}, {Lundkvist}, {Marcy}, {Metcalfe}, {Silva Aguirre}, \& {Winn}}]{Hubert+13}
{Huber}, D., {Carter}, J.~A., {Barbieri}, M., {et~al.} 2013, Science, 342, 331

\bibitem[{{Jensen} \& {Haugb{\o}lle}(2018)}]{Jensen+Haugboelle18}
{Jensen}, S.~S. \& {Haugb{\o}lle}, T. 2018, \mnras, 474, 1176

\bibitem[{{Jensen} {et~al.}(2023){Jensen}, {Spezzano}, {Caselli}, {Grassi}, \& {Haugb{\o}lle}}]{Jensen+23}
{Jensen}, S.~S., {Spezzano}, S., {Caselli}, P., {Grassi}, T., \& {Haugb{\o}lle}, T. 2023, \aap, 675, A34

\bibitem[{{Jermyn} {et~al.}(2023){Jermyn}, {Bauer}, {Schwab}, {Farmer}, {Ball}, {Bellinger}, {Dotter}, {Joyce}, {Marchant}, {Mombarg}, {Wolf}, {Sunny Wong}, {Cinquegrana}, {Farrell}, {Smolec}, {Thoul}, {Cantiello}, {Herwig}, {Toloza}, {Bildsten}, {Townsend}, \& {Timmes}}]{Jermyn+23}
{Jermyn}, A.~S., {Bauer}, E.~B., {Schwab}, J., {et~al.} 2023, \apjs, 265, 15

\bibitem[{{J{\o}rgensen} {et~al.}(2022){J{\o}rgensen}, {Kuruwita}, {Harsono}, {Haugb{\o}lle}, {Kristensen}, \& {Bergin}}]{Jorgensen+22}
{J{\o}rgensen}, J.~K., {Kuruwita}, R.~L., {Harsono}, D., {et~al.} 2022, \nat, 606, 272

\bibitem[{{Juvela}(2019)}]{Juvela2019}
{Juvela}, M. 2019, \aap, 622, A79

\bibitem[{{Juvela} {et~al.}(2020){Juvela}, {Neha}, {Mannfors}, {Saajasto}, {Ysard}, \& {Pelkonen}}]{Juvela2020_L1642}
{Juvela}, M., {Neha}, S., {Mannfors}, E., {et~al.} 2020, \aap, 643, A132

\bibitem[{{Kamiaka} {et~al.}(2019){Kamiaka}, {Benomar}, {Suto}, {Dai}, {Masuda}, \& {Winn}}]{Kamiaka+19}
{Kamiaka}, S., {Benomar}, O., {Suto}, Y., {et~al.} 2019, \aj, 157, 137

\bibitem[{{Kawai} {et~al.}(2024){Kawai}, {Narita}, {Fukui}, {Watanabe}, \& {Inaba}}]{Kawai+24}
{Kawai}, Y., {Narita}, N., {Fukui}, A., {Watanabe}, N., \& {Inaba}, S. 2024, \mnras, 528, 270

\bibitem[{{Kido} {et~al.}(2023){Kido}, {Takakuwa}, {Saigo}, {Ohashi}, {Tobin}, {J{\o}rgensen}, {Aikawa}, {Aso}, {Encalada}, {Flores}, {Gavino}, {de Gregorio-Monsalvo}, {Han}, {Hirano}, {Koch}, {Kwon}, {Lai}, {Lee}, {Lee}, {Li}, {Lin}, {Looney}, {Mori}, {Narayanan}, {Plunkett}, {Phuong}, {(Insa Choi)}, {Santamar{\'\i}a-Miranda}, {Sharma}, {Sheehan}, {Thieme}, {Tomida}, {van't Hoff}, {Williams}, {Yamato}, \& {Yen}}]{Kido+2023}
{Kido}, M., {Takakuwa}, S., {Saigo}, K., {et~al.} 2023, \apj, 953, 190

\bibitem[{{Kounkel} {et~al.}(2023){Kounkel}, {Stassun}, {Hillenbrand}, {Hern{\'a}ndez}, {Serna}, \& {Curtis}}]{Kounkel+23}
{Kounkel}, M., {Stassun}, K.~G., {Hillenbrand}, L.~A., {et~al.} 2023, \aj, 165, 182

\bibitem[{{Kuffmeier} {et~al.}(2021){Kuffmeier}, {Dullemond}, {Reissl}, \& {Goicovic}}]{Kuffmeier+21}
{Kuffmeier}, M., {Dullemond}, C.~P., {Reissl}, S., \& {Goicovic}, F.~G. 2021, \aap, 656, A161

\bibitem[{{Kuffmeier} {et~al.}(2018){Kuffmeier}, {Frimann}, {Jensen}, \& {Haugb{\o}lle}}]{Kuffmeier+18}
{Kuffmeier}, M., {Frimann}, S., {Jensen}, S.~S., \& {Haugb{\o}lle}, T. 2018, \mnras, 475, 2642

\bibitem[{{Kuffmeier} {et~al.}(2020){Kuffmeier}, {Goicovic}, \& {Dullemond}}]{Kuffmeier+20}
{Kuffmeier}, M., {Goicovic}, F.~G., \& {Dullemond}, C.~P. 2020, \aap, 633, A3

\bibitem[{{Kuffmeier} {et~al.}(2017){Kuffmeier}, {Haugb{\o}lle}, \& {Nordlund}}]{Kuffmeier+17}
{Kuffmeier}, M., {Haugb{\o}lle}, T., \& {Nordlund}, {\r{A}}. 2017, \apj, 846, 7

\bibitem[{{Kuffmeier} {et~al.}(2023){Kuffmeier}, {Jensen}, \& {Haugb{\o}lle}}]{Kuffmeier+23}
{Kuffmeier}, M., {Jensen}, S.~S., \& {Haugb{\o}lle}, T. 2023, European Physical Journal Plus, 138, 272

\bibitem[{{Kuffmeier} {et~al.}(2024){Kuffmeier}, {Pineda}, {Segura-Cox}, \& {Haugb{\o}lle}}]{Kuffmeier+24}
{Kuffmeier}, M., {Pineda}, J.~E., {Segura-Cox}, D., \& {Haugb{\o}lle}, T. 2024, \aap, 690, A297

\bibitem[{{Larson}(1981)}]{Larson81}
{Larson}, R.~B. 1981, MNRAS, 194, 809

\bibitem[{{Lee} {et~al.}(2021){Lee}, {Charnoz}, \& {Hennebelle}}]{Lee+21}
{Lee}, Y.-N., {Charnoz}, S., \& {Hennebelle}, P. 2021, \aap, 648, A101

\bibitem[{{Long} {et~al.}(2022){Long}, {Andrews}, {Rosotti}, {Harsono}, {Pinilla}, {Wilner}, {{\"O}berg}, {Teague}, {Trapman}, \& {Tabone}}]{Long+22}
{Long}, F., {Andrews}, S.~M., {Rosotti}, G., {et~al.} 2022, \apj, 931, 6

\bibitem[{{Louden} {et~al.}(2021){Louden}, {Winn}, {Petigura}, {Isaacson}, {Howard}, {Masuda}, {Albrecht}, \& {Kosiarek}}]{Louden+21}
{Louden}, E.~M., {Winn}, J.~N., {Petigura}, E.~A., {et~al.} 2021, \aj, 161, 68

\bibitem[{{Manara} {et~al.}(2023){Manara}, {Ansdell}, {Rosotti}, {Hughes}, {Armitage}, {Lodato}, \& {Williams}}]{Manara+23}
{Manara}, C.~F., {Ansdell}, M., {Rosotti}, G.~P., {et~al.} 2023, in Astronomical Society of the Pacific Conference Series, Vol. 534, Protostars and Planets VII, ed. S.~{Inutsuka}, Y.~{Aikawa}, T.~{Muto}, K.~{Tomida}, \& M.~{Tamura}, 539

\bibitem[{{Manara} {et~al.}(2018){Manara}, {Morbidelli}, \& {Guillot}}]{Manara+18}
{Manara}, C.~F., {Morbidelli}, A., \& {Guillot}, T. 2018, \aap, 618, L3

\bibitem[{{Manara} {et~al.}(2020){Manara}, {Natta}, {Rosotti}, {Alcal{\'a}}, {Nisini}, {Lodato}, {Testi}, {Pascucci}, {Hillenbrand}, {Carpenter}, {Scholz}, {Fedele}, {Frasca}, {Mulders}, {Rigliaco}, {Scardoni}, \& {Zari}}]{Manara+20}
{Manara}, C.~F., {Natta}, A., {Rosotti}, G.~P., {et~al.} 2020, \aap, 639, A58

\bibitem[{{Manara} {et~al.}(2017){Manara}, {Testi}, {Herczeg}, {Pascucci}, {Alcal{\'a}}, {Natta}, {Antoniucci}, {Fedele}, {Mulders}, {Henning}, {Mohanty}, {Prusti}, \& {Rigliaco}}]{Manara+17}
{Manara}, C.~F., {Testi}, L., {Herczeg}, G.~J., {et~al.} 2017, \aap, 604, A127

\bibitem[{{Manara} {et~al.}(2013){Manara}, {Testi}, {Rigliaco}, {Alcal{\'a}}, {Natta}, {Stelzer}, {Biazzo}, {Covino}, {Covino}, {Cupani}, {D'Elia}, \& {Randich}}]{Manara+13}
{Manara}, C.~F., {Testi}, L., {Rigliaco}, E., {et~al.} 2013, \aap, 551, A107

\bibitem[{{Marino} {et~al.}(2015){Marino}, {Perez}, \& {Casassus}}]{Marino+15}
{Marino}, S., {Perez}, S., \& {Casassus}, S. 2015, \apjl, 798, L44

\bibitem[{{Mathis} {et~al.}(1983){Mathis}, {Mezger}, \& {Panagia}}]{Mathis1983}
{Mathis}, J.~S., {Mezger}, P.~G., \& {Panagia}, N. 1983, \aap, 128, 212

\bibitem[{{Mauc{\'o}} {et~al.}(2023){Mauc{\'o}}, {Manara}, {Ansdell}, {Bettoni}, {Claes}, {Alcala}, {Miotello}, {Facchini}, {Haworth}, {Lodato}, \& {Williams}}]{Mauco+23}
{Mauc{\'o}}, K., {Manara}, C.~F., {Ansdell}, M., {et~al.} 2023, \aap, 679, A82

\bibitem[{{Moeckel} \& {Throop}(2009)}]{Moeckel+Throop09}
{Moeckel}, N. \& {Throop}, H.~B. 2009, \apj, 707, 268

\bibitem[{{Mulders} {et~al.}(2021){Mulders}, {Pascucci}, {Ciesla}, \& {Fernandes}}]{Mulders+21}
{Mulders}, G.~D., {Pascucci}, I., {Ciesla}, F.~J., \& {Fernandes}, R.~B. 2021, \apj, 920, 66

\bibitem[{{Ohashi} {et~al.}(2014){Ohashi}, {Saigo}, {Aso}, {Aikawa}, {Koyamatsu}, {Machida}, {Saito}, {Takahashi}, {Takakuwa}, {Tomida}, {Tomisaka}, \& {Yen}}]{Ohashi+14}
{Ohashi}, N., {Saigo}, K., {Aso}, Y., {et~al.} 2014, \apj, 796, 131

\bibitem[{{Okoda} {et~al.}(2021){Okoda}, {Oya}, {Francis}, {Johnstone}, {Inutsuka}, {Ceccarelli}, {Codella}, {Chandler}, {Sakai}, {Aikawa}, {Alves}, {Balucani}, {Bianchi}, {Bouvier}, {Caselli}, {Caux}, {Charnley}, {Choudhury}, {De Simone}, {Dulieu}, {Dur{\'a}n}, {Evans}, {Favre}, {Fedele}, {Feng}, {Fontani}, {Hama}, {Hanawa}, {Herbst}, {Hirota}, {Imai}, {Isella}, {J{\'\i}menez-Serra}, {Kahane}, {Lefloch}, {Loinard}, {L{\'o}pez-Sepulcre}, {Maud}, {Maureira}, {Menard}, {Mercimek}, {Miotello}, {Moellenbrock}, {Mori}, {Murillo}, {Nakatani}, {Nomura}, {Oba}, {O'Donoghue}, {Ohashi}, {Ospina-Zamudio}, {Pineda}, {Podio}, {Rimola}, {Sakai}, {Segura-Cox}, {Shirley}, {Svoboda}, {Taquet}, {Testi}, {Vastel}, {Viti}, {Watanabe}, {Watanabe}, {Witzel}, {Xue}, {Zhang}, {Zhao}, \& {Yamamoto}}]{Okoda+21}
{Okoda}, Y., {Oya}, Y., {Francis}, L., {et~al.} 2021, \apj, 910, 11

\bibitem[{{Padoan} {et~al.}(2014){Padoan}, {Haugb{\o}lle}, \& {Nordlund}}]{Padoan+14_luminosity}
{Padoan}, P., {Haugb{\o}lle}, T., \& {Nordlund}, {\AA}. 2014, \apj, 797, 32

\bibitem[{{Padoan} {et~al.}(2005){Padoan}, {Kritsuk}, {Norman}, \& {Nordlund}}]{Padoan+05_BH}
{Padoan}, P., {Kritsuk}, A., {Norman}, M.~L., \& {Nordlund}, {\AA}. 2005, \apjl, 622, L61

\bibitem[{{Padoan} {et~al.}(2020){Padoan}, {Pan}, {Juvela}, {Haugb{\o}lle}, \& {Nordlund}}]{Padoan+20massive}
{Padoan}, P., {Pan}, L., {Juvela}, M., {Haugb{\o}lle}, T., \& {Nordlund}, {\r{A}}. 2020, \apj, 900, 82

\bibitem[{{Padoan} {et~al.}(2025){Padoan}, {Pan}, {Pelkonen}, {Haugb{\o}lle}, \& {Nordlund}}]{Padoan+24_angular}
{Padoan}, P., {Pan}, L., {Pelkonen}, V.-M., {Haugb{\o}lle}, T., \& {Nordlund}, {\r{A}}. 2025, Nature Astronomy (in press), pre-print available at arXiv:2405.07334

\bibitem[{{Paxton} {et~al.}(2011){Paxton}, {Bildsten}, {Dotter}, {Herwig}, {Lesaffre}, \& {Timmes}}]{Paxton+11}
{Paxton}, B., {Bildsten}, L., {Dotter}, A., {et~al.} 2011, \apjs, 192, 3

\bibitem[{{Paxton} {et~al.}(2013){Paxton}, {Cantiello}, {Arras}, {Bildsten}, {Brown}, {Dotter}, {Mankovich}, {Montgomery}, {Stello}, {Timmes}, \& {Townsend}}]{Paxton+13}
{Paxton}, B., {Cantiello}, M., {Arras}, P., {et~al.} 2013, \apjs, 208, 4

\bibitem[{{Paxton} {et~al.}(2015){Paxton}, {Marchant}, {Schwab}, {Bauer}, {Bildsten}, {Cantiello}, {Dessart}, {Farmer}, {Hu}, {Langer}, {Townsend}, {Townsley}, \& {Timmes}}]{Paxton+15}
{Paxton}, B., {Marchant}, P., {Schwab}, J., {et~al.} 2015, \apjs, 220, 15

\bibitem[{{Paxton} {et~al.}(2018){Paxton}, {Schwab}, {Bauer}, {Bildsten}, {Blinnikov}, {Duffell}, {Farmer}, {Goldberg}, {Marchant}, {Sorokina}, {Thoul}, {Townsend}, \& {Timmes}}]{Paxton+18}
{Paxton}, B., {Schwab}, J., {Bauer}, E.~B., {et~al.} 2018, \apjs, 234, 34

\bibitem[{{Paxton} {et~al.}(2019){Paxton}, {Smolec}, {Schwab}, {Gautschy}, {Bildsten}, {Cantiello}, {Dotter}, {Farmer}, {Goldberg}, {Jermyn}, {Kanbur}, {Marchant}, {Thoul}, {Townsend}, {Wolf}, {Zhang}, \& {Timmes}}]{Paxton+19}
{Paxton}, B., {Smolec}, R., {Schwab}, J., {et~al.} 2019, \apjs, 243, 10

\bibitem[{{Pelkonen} {et~al.}(2021){Pelkonen}, {Padoan}, {Haugb{\o}lle}, \& {Nordlund}}]{Pelkonen+21}
{Pelkonen}, V.~M., {Padoan}, P., {Haugb{\o}lle}, T., \& {Nordlund}, {\r{A}}. 2021, \mnras, 504, 1219

\bibitem[{{Pineda} {et~al.}(2020){Pineda}, {Segura-Cox}, {Caselli}, {Cunningham}, {Zhao}, {Schmiedeke}, {Maureira}, \& {Neri}}]{Pineda+20}
{Pineda}, J.~E., {Segura-Cox}, D., {Caselli}, P., {et~al.} 2020, Nature Astronomy, 4, 1158

\bibitem[{{Pinilla} {et~al.}(2018){Pinilla}, {Benisty}, {de Boer}, {Manara}, {Bouvier}, {Dominik}, {Ginski}, {Loomis}, \& {Sicilia Aguilar}}]{Pinilla+18}
{Pinilla}, P., {Benisty}, M., {de Boer}, J., {et~al.} 2018, \apj, 868, 85

\bibitem[{{Sai} {et~al.}(2020){Sai}, {Ohashi}, {Saigo}, {Matsumoto}, {Aso}, {Takakuwa}, {Aikawa}, {Kurose}, {Yen}, {Tomisaka}, {Tomida}, \& {Machida}}]{Sai+20}
{Sai}, J., {Ohashi}, N., {Saigo}, K., {et~al.} 2020, \apj, 893, 51

\bibitem[{{Sai} {et~al.}(2024){Sai}, {Yen}, {Machida}, {Ohashi}, {Aso}, {Maury}, \& {Maret}}]{Sai+24}
{Sai}, J., {Yen}, H.-W., {Machida}, M.~N., {et~al.} 2024, \apj, 966, 192

\bibitem[{{Salpeter}(1955)}]{Salpeter55}
{Salpeter}, E.~E. 1955, ApJ, 121, 161

\bibitem[{{Sanchis-Ojeda} {et~al.}(2013){Sanchis-Ojeda}, {Winn}, {Marcy}, {Howard}, {Isaacson}, {Johnson}, {Torres}, {Albrecht}, {Campante}, {Chaplin}, {Davies}, {Lund}, {Carter}, {Dawson}, {Buchhave}, {Everett}, {Fischer}, {Geary}, {Gilliland}, {Horch}, {Howell}, \& {Latham}}]{Sanchis-Ojeda+13}
{Sanchis-Ojeda}, R., {Winn}, J.~N., {Marcy}, G.~W., {et~al.} 2013, \apj, 775, 54

\bibitem[{Solomon {et~al.}(1987)Solomon, Rivolo, Barrett, \& Yahil}]{Solomon+87}
Solomon, P.~M., Rivolo, A.~R., Barrett, J.~W., \& Yahil, A.~M. 1987, ApJ, 319, 730

\bibitem[{{Stapper} {et~al.}(2022){Stapper}, {Hogerheijde}, {van Dishoeck}, \& {Mentel}}]{Stapper+22}
{Stapper}, L.~M., {Hogerheijde}, M.~R., {van Dishoeck}, E.~F., \& {Mentel}, R. 2022, \aap, 658, A112

\bibitem[{{Stef{\'a}nsson} {et~al.}(2023){Stef{\'a}nsson}, {Mahadevan}, {Miguel}, {Robertson}, {Delamer}, {Kanodia}, {Ca{\~n}as}, {Winn}, {Ninan}, {Terrien}, {Holcomb}, {Ford}, {Zawadzki}, {Bowler}, {Bender}, {Cochran}, {Diddams}, {Endl}, {Fredrick}, {Halverson}, {Hearty}, {Hill}, {Lin}, {Metcalf}, {Monson}, {Ramsey}, {Roy}, {Schwab}, {Wright}, \& {Zeimann}}]{Stefansson+23}
{Stef{\'a}nsson}, G., {Mahadevan}, S., {Miguel}, Y., {et~al.} 2023, Science, 382, 1031

\bibitem[{{Steinacker} {et~al.}(2010){Steinacker}, {Pagani}, {Bacmann}, \& {Guieu}}]{Steinacker2010}
{Steinacker}, J., {Pagani}, L., {Bacmann}, A., \& {Guieu}, S. 2010, \aap, 511, A9

\bibitem[{{Testi} {et~al.}(2022){Testi}, {Natta}, {Manara}, {de Gregorio Monsalvo}, {Lodato}, {Lopez}, {Muzic}, {Pascucci}, {Sanchis}, {Miranda}, {Scholz}, {De Simone}, \& {Williams}}]{Testi+22}
{Testi}, L., {Natta}, A., {Manara}, C.~F., {et~al.} 2022, \aap, 663, A98

\bibitem[{{Teyssier}(2007)}]{Teyssier07}
{Teyssier}, R. 2007, Geophysical and Astrophysical Fluid Dynamics, 101, 199

\bibitem[{{Thieme} {et~al.}(2022){Thieme}, {Lai}, {Lin}, {Cheong}, {Lee}, {Yen}, {Li}, {Lam}, \& {Zhao}}]{Thieme+22}
{Thieme}, T.~J., {Lai}, S.-P., {Lin}, S.-J., {et~al.} 2022, \apj, 925, 32

\bibitem[{{Throop} \& {Bally}(2008)}]{Throop+Bally08}
{Throop}, H.~B. \& {Bally}, J. 2008, \aj, 135, 2380

\bibitem[{{Valdivia-Mena} {et~al.}(2024){Valdivia-Mena}, {Pineda}, {Caselli}, {Segura-Cox}, {Schmiedeke}, {Spezzano}, {Offner}, {Ivlev}, {Kuffmeier}, {Cunningham}, {Neri}, \& {Maureira}}]{Valdivia+24}
{Valdivia-Mena}, M.~T., {Pineda}, J.~E., {Caselli}, P., {et~al.} 2024, \aap, 687, A71

\bibitem[{{Valdivia-Mena} {et~al.}(2022){Valdivia-Mena}, {Pineda}, {Segura-Cox}, {Caselli}, {Neri}, {L{\'o}pez-Sepulcre}, {Cunningham}, {Bouscasse}, {Semenov}, {Henning}, {Pi{\'e}tu}, {Chapillon}, {Dutrey}, {Fuente}, {Guilloteau}, {Hsieh}, {Jim{\'e}nez-Serra}, {Marino}, {Maureira}, {Smirnov-Pinchukov}, {Tafalla}, \& {Zhao}}]{Valdivia+22}
{Valdivia-Mena}, M.~T., {Pineda}, J.~E., {Segura-Cox}, D.~M., {et~al.} 2022, \aap, 667, A12

\bibitem[{{Weingartner} \& {Draine}(2001)}]{Weingartner2001}
{Weingartner}, J.~C. \& {Draine}, B.~T. 2001, \apj, 548, 296

\bibitem[{{Yamato} {et~al.}(2023){Yamato}, {Aikawa}, {Ohashi}, {Tobin}, {J{\o}rgensen}, {Takakuwa}, {Aso}, {Sai}, {Flores}, {de Gregorio-Monsalvo}, {Hirano}, {Han}, {Kido}, {Koch}, {Kwon}, {Lai}, {Lee}, {Lee}, {Li}, {Lin}, {Looney}, {Mori}, {Narayanan}, {Phuong}, {Saigo}, {Santamar{\'\i}a-Miranda}, {Sharma}, {Thieme}, {Tomida}, {van't Hoff}, \& {Yen}}]{Yamato+23}
{Yamato}, Y., {Aikawa}, Y., {Ohashi}, N., {et~al.} 2023, \apj, 951, 11

\bibitem[{{Yen} {et~al.}(2019){Yen}, {Gu}, {Hirano}, {Koch}, {Lee}, {Liu}, \& {Takakuwa}}]{Yen+19}
{Yen}, H.-W., {Gu}, P.-G., {Hirano}, N., {et~al.} 2019, \apj, 880, 69

\bibitem[{{Yen} \& {Lee}(2024)}]{YenLee24}
{Yen}, H.-W. \& {Lee}, Y.-N. 2024, \apjl, 972, L27

\bibitem[{{Yen} {et~al.}(2024){Yen}, {Williams}, {Sai}, {Koch}, {Han}, {J{\o}rgensen}, {Kwon}, {Lee}, {Li}, {Looney}, {Narang}, {Ohashi}, {Takakuwa}, {Tobin}, {de Gregorio-Monsalvo}, {Lai}, {Lee}, \& {Tomida}}]{Yen+24}
{Yen}, H.-W., {Williams}, J.~P., {Sai}, J., {et~al.} 2024, \apj, 969, 125

\end{thebibliography}




\end{document}